\documentclass[a4paper]{article}
\usepackage{fullpage}
\usepackage{xcolor}
\usepackage{hyperref}

\author{Hideo Bannai \and Juha K\"{a}rkk\"{a}inen \and Dominik K\"{o}ppl \and Marcin~Pi\c{a}tkowski}

\usepackage[utf8]{inputenc}
\usepackage{adjustbox}

\usepackage{savesym}
\savesymbol{iint}
\savesymbol{iiint}
\savesymbol{iiiint}
\savesymbol{idotsint}
\usepackage{amsmath}
\restoresymbol{AMS}{iint}
\restoresymbol{AMS}{iiint}
\restoresymbol{AMS}{iiiint}
\restoresymbol{AMS}{idotsint}

\usepackage{placeins}

\usepackage{amssymb}
\usepackage{amsthm}

\usepackage{booktabs}

\usepackage[cref,color,theorems]{nikis}

\newcommand*{\instancename}[1]{\ensuremath{\mathsf{#1}}} \newcommand*{\functionname}[1]{{\ensuremath{\renewcommand{\rmdefault}{ptm}\fontfamily{ppl}\selectfont\textrm{\textup{#1}}}}}

\definecolor{teigiColor}{HTML}{5700B5}
\newcommand*{\teigi}[1]{\emph{\color{teigiColor}#1}}
\newcommand*{\typeL}{\texttt{L}}
\newcommand*{\typeS}{\texttt{S}}
\newcommand*{\typeHoshi}{\ensuremath{\texttt{S}^*}}

\newcommand*{\SA}{\instancename{SA}}
\newcommand*{\CSA}{\ensuremath{\instancename{SA}_{\circ}}}
\newcommand*{\BBWT}{\instancename{BBWT}}
\newcommand*{\BWT}{\instancename{BWT}}

\newcommand*{\ibeg}[1]{\ensuremath{\mathsf{b}(#1)}}\newcommand*{\iend}[1]{\ensuremath{\mathsf{e}(#1)}}

\newcommand*{\RunningExample}{cbbcacbbcadacbadacba}

\newcommand*{\GenOrder}{\ensuremath{\prec_{\textup{gen}}}}
\newcommand*{\OmegaOrder}{\ensuremath{\prec_\omega}}
\newcommand*{\OmegaOrderSucc}{\ensuremath{\succ_\omega}}
\newcommand*{\LMSOrder}{\ensuremath{\prec_{\textup{LMS}}}}
\newcommand*{\LexOrder}{\ensuremath{\prec_{\textup{lex}}}}
\newcommand*{\LexOrderSucc}{\ensuremath{\succ_{\textup{lex}}}}
\newcommand*{\LexOrderSuccEq}{\ensuremath{\succeq_{\textup{lex}}}}

\newcommand*{\conj}[2][]{\ensuremath{\functionname{conj}_{#1}(#2)}}

\usepackage{tikz}
\usetikzlibrary{matrix,fit,positioning}
\tikzstyle{InductionMatrix} = [matrix of nodes
    , nodes in empty cells
	, column sep=0.5\pgflinewidth
	, row sep=0.5mm
	, row 1/.style={nodes={font=\tiny,minimum width=2em,fill=none, minimum size=0mm, text height=0.33em}}
	, row 2/.style={nodes={minimum width=1.6em}}
	, row 5/.style={nodes={minimum height=4em}}
	, row 8/.style={nodes={font=\ttfamily}}
	, column 21/.style={nodes={font=\tiny}}
	]

\tikzstyle{LyndonBlock} = [inner sep=0, inner xsep=5, draw=solarizedViolet,rounded corners=4pt]
\tikzstyle{LMSmatrix} = [matrix of nodes
    , nodes in empty cells
, row sep=0.5mm
	, inner sep=0.25em,
	, row 2/.style={nodes={font=\ttfamily,scale=0.9, fill=none, minimum size=0mm, text height=0.5em}}
	, row 3/.style={nodes={font=\tiny,fill=none, minimum size=0mm, text height=0.33em}}
	, row 1/.style={nodes={font=\tiny,fill=none, minimum size=0mm, text height=0.33em}}
	]

	\newcommand*{\LMSbracket}[2]{\begin{scope}[transform canvas={yshift=-0.5em,xshift=-0.3em}]
	  \draw [color=solarizedBlue] ([yshift=0.3em,xshift=+0.2em]text-3-#1.south) -- ([xshift=+0.2em]text-3-#1.south) -- (text-3-#2.south) --  ([yshift=0.3em]text-3-#2.south)  {};
	\end{scope}
}
\tikzstyle{InductionMatrixSubstring} = [InductionMatrix,
, row 3/.style={nodes={text height=1em}}
, row 4/.style={nodes={text height=2em}}
, row 5/.style={nodes={text height=1em}}
, row 6/.style={nodes={text height=1em}}
, row 7/.style={nodes={text height=0.3em}}
	]

\newcounter{PfeilCount}
	\newcommand{\Pfeil}[5]{\pgfmathsetmacro{\redValue}{(mod(\thePfeilCount * 172, 256))}
	  \pgfmathsetmacro{\greenValue}{(mod(\thePfeilCount * 172 + 123, 256))}
	  \pgfmathsetmacro{\blueValue}{(mod(\thePfeilCount * 171 + 181, 256))}
	  \draw [color={rgb:red,\redValue;green,\greenValue;blue,\blueValue}, ->] ([xshift=0.2em]sa-#1-#2.south) -- ([xshift=0.2em,yshift=-#5]sa-#3-#2.south)  -- ([xshift=-0.2em,yshift=-#5]sa-#3-#4.south) -- ([xshift=-0.2em]sa-#3-#4.south);
	  \stepcounter{PfeilCount}
	}

\newcommand*{\PatternLegendS}[1]{\tikz[baseline=.35ex]{\draw[draw,#1] (0, 0) rectangle (.35, .35); }}
\newcommand*{\PatternLegend}[1]{(\scalebox{0.7}{\PatternLegendS{#1}})}
\usetikzlibrary{patterns}

\title{Constructing the Bijective and the Extended Burrows--Wheeler Transform in Linear Time}

\begin{document}

\date{}
\maketitle

\begin{abstract}
The Burrows--Wheeler transform (BWT) is a permutation whose applications
are prevalent in data compression and text indexing.
The \emph{bijective BWT} (BBWT) is a bijective variant of it.
Although it is known that the BWT can be constructed in linear time for integer alphabets by using a linear time suffix array construction algorithm, it was up to now only conjectured that the BBWT can also be constructed in linear time.
We confirm this conjecture in the word RAM model by proposing a construction algorithm that is based on SAIS, 
improving the best known result of \Oh{n \lg n /\lg \lg n} time to linear.
Since we can reduce the problem of constructing the extended BWT to constructing the BBWT in linear time, we obtain a linear-time algorithm computing the extended BWT at the same time.
\end{abstract}
\renewcommand*{\subparagraph}[2]{\textbf{#2.~}}

\section{Introduction}

The Burrows--Wheeler transform (BWT)~\cite{burrows94bwt}
is a transformation permuting the characters of a given string~$T\texttt{\$}$,
where \texttt{\$} is a character that is strictly smaller than all characters occurring it~$T$.
The $i$-th entry of the BWT of~$T\texttt{\$}$ is the character preceding the $i$-th lexicographically smallest suffix of~$T\texttt{\$}$,
or \texttt{\$} if this suffix is $T\texttt{\$}$ itself.
Strictly speaking, the BWT is not a bijection since its output contains \texttt{\$} at an arbitrary position while 
it requests the input~$T$ to have \texttt{\$} as a delimiter at its end in order to restore~$T$.
A variant, called the bijective BWT~\cite{kufleitner09bwt,gil12bbwt}, is a \emph{bijective} transformation, 
which does not require the artificial delimiter~\texttt{\$}.
It is based on the Lyndon factorization~\cite{chen58lyndon} of~$T$.
In this variant, the output consists of the last characters of the lexicographically sorted
cyclic rotations of all factors composing the Lyndon factorization of~$T$.

In the following, we call the BWT \teigi{traditional} to ease the distinguishability of both transformations.
It is well known that the traditional BWT has many applications in data compression~\cite{adjeroh08bwt} and text indexing~\cite{ferragina00fmindex,ferragina05index,gagie18bwt}.
Recently, such a text index was adapted to work with the bijective BWT~\cite{bannai19bbwt}.

\subparagraph*{Related Work} 
In what follows, we review the traditional BWT construction via suffix arrays, and some algorithms computing the BBWT or the extended BWT\@.
For the complexity analysis, we take a text~$T$ of length~$n$ whose characters are drawn from a polynomial bounded integer alphabet~$\{1,\ldots,n^{\Oh{1}}\}$.
Let us start with the traditional BWT, which we can construct thanks to linear time suffix array construction algorithms~\cite{nong11sais,karkkainen06dc3}
in linear time.
That is because the traditional BWT, denoted by $\BWT[1..n]$, is determined by $\BWT[i] = T[\SA[i]-1]$ for $\SA[i] > 1$ and $\BWT[i] = T[n]$ for $\SA[i] =1$.
Considering the bijective BWT, 
{Gil and Scott}~\cite{gil12bbwt} postulated that it can be built in linear time,
but did not give a construction algorithm. 
It is clear that the time is upper bounded by the total length of all conjugates~\cite[after Example~9]{mantaci07ebwt},
which is \Oh{n^2}.
In the same paper, {Mantaci et al.}~\cite{mantaci07ebwt} also introduced the \teigi{extended} BWT, a generalization of the BBWT in that it is a BWT based on a set~$\mathcal{S}$ of \teigi{primitive} strings, i.e., strings that are not periodic.
{Hon et al.}~\cite{hon11circular} provided an algorithm building the extended BWT in \Oh{n \lg n} time.
Their idea is to construct the \teigi{circular suffix array} $\CSA{}$ such that the $i$-th position of the extended BWT is given by 
$T[\CSA[i]-1]$, where $T$ is the concatenation of all strings in $\mathcal{S}$.
{Bonomo et al.}~\cite{bonomo14sorting} presented the most recent algorithm building the bijective BWT online in $\Oh{n \lg n /\lg \lg n}$ time.
In~\cite[Sect.~6]{bonomo14sorting}, they also gave a linear time reduction from computing the extended BWT to computing the BBWT\@.
Knowing that an irreducible word has exactly one conjugate being a Lyndon word,
the reduction is done by exchanging each element of the set of irreducible strings~$\mathcal{S}$ by the conjugate being a Lyndon word, 
and concatenating these Lyndon words after sorting them in descending order.
Consequently, a linear-time BBWT construction algorithm can be used to compute the extended BWT in linear time.

On the practical side, we are aware of the work of 
Branden Brown\footnote{\url{https://github.com/zephyrtronium/bwst}},
Yuta Mori in his OpenBWT library\footnote{\url{https://web.archive.org/web/20170306035431/https://encode.ru/attachment.php?attachmentid=959&d=1249146089}},
and of Neal Burns\footnote{\texttt{https://github.com/NealB/Bijective-BWT}}.
While the first is a naive but easily understandable implementation calling a general sorting algorithm on all conjugates to directly compute the BBWT,
the second seems to be an adaptation of the suffix array -- induced sorting (SAIS) algorithm~\cite{nong11sais} to induce the BBWT\@.
The last one takes an already computed suffix array~\SA{} as input, and modifies \SA{} such that reading the characters $T[\SA[i]-1]$ gives the BBWT\@.
For that, this algorithm shifts entries in \SA{} to the right until they fit. 
Hence, the running time is based on the lengths of these shifts, which can be \Oh{n^2}, but seem to be negligible in practice for common texts.

\subparagraph*{Our Result}
In this article, we present a linear time algorithm computing the BBWT in the word RAM model.
The main idea is to adapt SAIS to compute the circular suffix array of the Lyndon factors.
We obtain linear running time by exploiting some facts based on the nature of the Lyndon factorization.

\section{Preliminaries}
Our computational model is the word RAM model with word size \Om{\lg n}.
Accessing a word costs \Oh{1} time.
In this article, we study strings on an \emph{integer} alphabet $\Sigma = \{1,\ldots,\sigma\}$ with size $\sigma = n^{\Oh{1}}$.

\block{Strings}
We call an element $T \in \Sigma^*$ a \teigi{string}.
Its length is denoted by $|T|$.
Given an integer~$j \in [1..|T|]$, we access the $j$-th character of~$T$ with~$T[j]$.
Given a string $T \in \Sigma^*$, we denote with $T^k$ that we concatenate $k$ times the string~$T$.
When $T$ is represented by the concatenation of $X, Y, Z \in \Sigma^*$, i.e., $T = \textit{XYZ}$,
then $X$, $Y$, and $Z$ are called a \teigi{prefix}, \teigi{substring}, and \teigi{suffix} of $T$, respectively.
A prefix $X$, substring~$Y$, or suffix~$Z$ is called \teigi{proper} if $X \neq T$, $Y \neq T$, or $Z \neq T$, respectively.
A proper prefix~$X$ of~$T$ is called a \teigi{border} of~$T$ if it is also a suffix of~$T$.
$T$ is called \teigi{border-free} if it has no border.
For two integers~$i$ and~$j$ with $1 \leq i \leq j \leq |T|$, let $T[i..j]$ denote
the substring of $T$ that begins at position~$i$ and ends at position~$j$ in~$T$.
If $i > j$, then $T[i..j]$ is the empty string.
In particular, the suffix starting at position~$j$ of~$T$ is denoted with $T[j..n]$.
A string~$T$ is called \teigi{primitive} if it cannot be written as $T = S^k$ for a string~$S \in \Sigma^+$ and $k \ge 2$.

\block{Orders on Strings}
We denote the \teigi{lexicographic order} with $\LexOrder$.
Given two strings $S$ and $T$, then $S \LexOrder T$ if $S$ is a proper prefix of $T$ or there exists an integer~$\ell$ with
$1 \le \ell \le \min(|S|,|T|)$ such that $S[1..\ell-1]=T[1..\ell-1]$ and $S[\ell] < T[\ell]$.
We write $S \OmegaOrder{} T$ if the infinite concatenation $S^\omega := {SSS} \cdots$ is lexicographically smaller than 
$T^\omega := {TTT}\cdots$. 
For instance, $\texttt{ab} \LexOrder \texttt{aba}$ but $\texttt{aba} \OmegaOrder \texttt{ab}$.
The relation $\OmegaOrder$ induces an order on the set of \emph{primitive} strings\footnote{The order cannot be generalized to strings in general since $\texttt{a} \neq \texttt{aa}$ but neither $\texttt{a} \OmegaOrder \texttt{aa}$ nor $\texttt{aa} \OmegaOrder \texttt{a}$ holds.},
which we call \teigi{$\OmegaOrder$-order}.

  \begin{figure}
    \centering{\scalebox{0.95}{\adjustbox{valign=t}{\rotatebox{90}{$T=$
      \texttt{\RunningExample{}}
  }}
  \hspace{0.5em}
      \adjustbox{valign=t}{\rotatebox{90}{\color{gray}
	$\downarrow$
      Lyndon Factorization
	$\downarrow$
  }}
  \hspace{0.5em}
      \adjustbox{valign=t}{\rotatebox{90}{\texttt{c} $\mid$
  \texttt{bbc} $\mid$
  \texttt{acbbcad} $\mid$
  \texttt{acbad} $\mid$
  \texttt{acb}$ \mid$
  \texttt{a}
  }}
  \hspace{0.5em}
      \adjustbox{valign=t}{\rotatebox{90}{\color{gray}
	$\downarrow$
	Collect the conjugates of all Lyndon factors
	$\downarrow$
  }}
  \hspace{0.5em}
      \begin{minipage}[t]{5em}
	\texttt{\color{solarizedYellow}c} \\
	\texttt{\color{solarizedYellow}bbc} \\
	\texttt{bcb} \\
	\texttt{cbb} \\
	\texttt{\color{solarizedYellow}acbbcad} \\
	\texttt{cbbcada} \\
	\texttt{bbcadac} \\
	\texttt{bcadacb} \\
	\texttt{cadacbb} \\
	\texttt{adacbbc} \\
	\texttt{dacbbca} \\
	\texttt{\color{solarizedYellow}acbad} \\
	\texttt{cbada} \\
	\texttt{badac} \\
	\texttt{adacb} \\
	\texttt{dacba} \\
	\texttt{\color{solarizedYellow}acb} \\
	\texttt{cba} \\
	\texttt{bac} \\
	\texttt{\color{solarizedYellow}a} \\
      \end{minipage}
      \adjustbox{valign=t}{\rotatebox{90}{\color{gray}
	$\downarrow$
	Sort them in $\OmegaOrder$-order
	$\downarrow$
  }}
  \hspace{0.5em}
      \begin{minipage}[t]{5em}
	\texttt{\color{solarizedYellow}a} \\
	\texttt{\color{solarizedYellow}acb} \\
	\texttt{\color{solarizedYellow}acbad} \\
	\texttt{\color{solarizedYellow}acbbcad} \\
	\texttt{adacb} \\
	\texttt{adacbbc} \\
	\texttt{bac} \\
	\texttt{badac} \\
	\texttt{bbcadac} \\
	\texttt{\color{solarizedYellow}bbc} \\
	\texttt{bcadacb} \\
	\texttt{bcb} \\
	\texttt{cadacbb} \\
	\texttt{cba} \\
	\texttt{cbada} \\
	\texttt{cbbcada} \\
	\texttt{cbb} \\
	\texttt{\color{solarizedYellow}c} \\
	\texttt{dacba} \\
	\texttt{dacbbca}
      \end{minipage}
   \hspace{0.5em}
      \adjustbox{valign=t}{\rotatebox{90}{\color{gray}$\downarrow$
	Set $\BBWT[i]$ to last character of $i$-th entry
	$\downarrow$
  }}
   \hspace{0.5em}
       \begin{minipage}[t]{1.5em}
	\texttt{\color{solarizedYellow}a} \\
	\texttt{\color{solarizedYellow}b} \\
	\texttt{\color{solarizedYellow}d} \\
	\texttt{\color{solarizedYellow}d} \\
	\texttt{b} \\
	\texttt{c} \\
	\texttt{c} \\
	\texttt{c} \\
	\texttt{c} \\
	\texttt{\color{solarizedYellow}c} \\
	\texttt{b} \\
	\texttt{b} \\
	\texttt{b} \\
	\texttt{a} \\
	\texttt{a} \\
	\texttt{a} \\
	\texttt{b} \\
	\texttt{\color{solarizedYellow}c} \\
	\texttt{a} \\
	\texttt{a}
       \end{minipage}
      \adjustbox{valign=t}{\rotatebox{90}{\color{gray}
	The corresponding starting position in the text:
  }}
   \hspace{0.5em}
       \begin{minipage}[t]{1.5em}
	\texttt{\color{solarizedYellow}20} \\
	\texttt{\color{solarizedYellow}17} \\
	\texttt{\color{solarizedYellow}12} \\
	\texttt{\color{solarizedYellow}\phantom{.}5} \\
	\texttt{15} \\
	\texttt{10} \\
	\texttt{19} \\
	\texttt{14} \\
	\texttt{\phantom{.}7} \\
	\texttt{\color{solarizedYellow}\phantom{.}2} \\
	\texttt{\phantom{.}8} \\
	\texttt{\phantom{.}3} \\
	\texttt{\phantom{.}9} \\
	\texttt{18} \\
	\texttt{13} \\
	\texttt{\phantom{.}6} \\
	\texttt{\phantom{.}4} \\
	\texttt{\color{solarizedYellow}\phantom{.}1} \\
	\texttt{16} \\
	\texttt{11}
       \end{minipage}
}}
\caption{Constructing \BBWT{} of $T = \texttt{\RunningExample{}}$. 
The Lyndon factors are highlighted~\protect\PatternLegend{fill=solarizedYellow}.
Reading the characters of the penultimate column top-down yields \BBWT{}.
The last column shows in its $i$-th row the starting position of the $i$-th smallest conjugate of a Lyndon factor in the text.
It is the circular suffix array studied later in \cref{secComposedString}.
Note that $\texttt{cbb} \LexOrder \texttt{cbbcada}$, but $\texttt{cbbcada} \OmegaOrder \texttt{cbb}$.
}
    \label{figComputingBBWT}
  \end{figure}

\block{Lyndon Words}
Given a primitive string~$T = T[1..n]$, its $i$-th \teigi{conjugate} \conj[i]{T} is defined as $T[i+1..n]T[1..i]$ for an integer $i \in [0..n-1]$.
Since $T$ is primitive, all its conjugates are distinct.
We say that $T$ and every one of its conjugates belongs to the \teigi{conjugate class} $\conj{T} := \{ \conj[0]{T}, \ldots, \conj[n-1]{T} \}$.
If a conjugate class contains \emph{exactly} one conjugate that is lexicographically smaller than all other conjugates,
then this conjugate is called a \teigi{Lyndon word}~\cite{lyndon54}.
Equivalently, a string $T$ is said to be a Lyndon word if and only if $T \prec S$ for every proper suffix $S$ of $T$.
A consequence is that a Lyndon word is \teigi{border-free}.

The \teigi{Lyndon factorization}~\cite{chen58lyndon} of $T\in\Sigma^+$ 
is the unique factorization of $T$ into a sequence of Lyndon words~$F_1 \cdots F_z$,
where (a) each $F_x\in\Sigma^+$ is a Lyndon word, and (b) $F_x \LexOrderSuccEq F_{x+1}$ for each $x \in [1..z)$.

\begin{lemma}[{\cite[Algo.\ 2.1]{duval83lyndon}}]\label{lem:LyndonFactorizationLinearTime}
	The Lyndon factorization of a string can be computed in linear time.
\end{lemma}

Each Lyndon word $F_x$ for $x \in [1..z]$ is called a \teigi{Lyndon factor}.
For what follows, we fix a string $T[1..n]$ over an alphabet $\Sigma$ of size~$\sigma$.
We use the string $T := \texttt{\RunningExample}$ as our running example.
Its Lyndon factorization is $\texttt{c}, \texttt{bbc}, \texttt{acbbcad}, \texttt{acbad}, \texttt{acb}, \texttt{a}$.

\block{Bijective Burrows--Wheeler Transform}
We denote the bijective BWT of~$T$ by \BBWT{}, where $\BBWT[i]$ is the last character of the $i$-th string in the list 
storing the conjugates of all Lyndon factors $F_1,\ldots,F_z$ of~$T$ sorted with respect to \OmegaOrder{}. 
\Cref{figComputingBBWT} shows the BBWT of our running example.

\begin{figure}
    \centering{\begin{tikzpicture}
\matrix[LMSmatrix] (text) {1 & 2  & 3 & 4 & 5  & 6 & 7  & 8 & 9 & 10 & 11 & 12 & 13 & 14 & 15 & 16 & 17 & 18 & 19 & 20   \\
c   & b  & b & c & a  & c & b  & b & c & a  & d  & a  & c  & b  & a  & d  & a  & c  & b  & a \\
L  & S* & S & L & S* & L & S* & S & L & S* & L  & S*  & L  & L  & S* & L  & S* & L  & L  & S*  \\
	};
	\node [left=0cm of text-2-1] {$T = $};

\LMSbracket{2}{5}
	
\LMSbracket{5}{7}
\LMSbracket{7}{10}
\LMSbracket{10}{12}
\LMSbracket{12}{15}
	\LMSbracket{15}{17}
	\LMSbracket{17}{20}
	\LMSbracket{20}{20}
\end{tikzpicture}
\hfill
\begin{tikzpicture}
\matrix[LMSmatrix] (text) {1 & 2  & 3 & 4 & 5  & 6 & 7 & 8\\
  E & C  & E & D & B  & D & B & A  \\
  L & S* & L & L & S* & L & L & S*\\
	};
	\node [left=0cm of text-2-1] {$T^{(1)} = $};

	\LMSbracket{2}{5}

	\LMSbracket{5}{8}
	\LMSbracket{8}{8}

\end{tikzpicture}
}\caption{Splitting $T$ and $T^{(1)}$ into LMS substrings. 
	The rectangular brackets below the types represent the LMS substrings.
	$T^{(1)}$ is $T$ after the replacement of its LMS substrings with their corresponding ranks defined in \cref{secExample} and on the left of \cref{figExRankingSAIS}.
    }
    \label{figExLMSSAIS}

\end{figure}
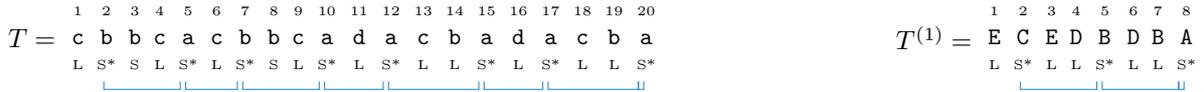

\section{Reviewing SAIS}\label{secSAIS}
Our idea is to adapt SAIS to compute $\CSA$ instead of the suffix array.
To explain this adaptation, we briefly review SAIS\@.
First, SAIS assigns each suffix a type, which is either~$\typeL$ or~$\typeS$:
\begin{itemize}
  \item $T[i..|T|]$ is an \typeL{} suffix if $T[i..|T|] \LexOrderSucc T[i+1..|T|]$, or
\item $T[i..|T|]$ is an \typeS{} suffix otherwise, i.e., $T[i..|T|] \LexOrder T[i+1..|T|]$,
\end{itemize}
where we stipulate that $T[|T|]$ is always type~$\typeS$.
Since it is not possible that $T[i..|T|] = T[i+1..|T|]$, SAIS assigns each suffix a type.
An \typeS{} suffix $T[i..|T|]$ is additionally an \typeHoshi{} suffix (also called LMS suffix in~\cite{nong11sais})
if $T[i-1..|T|]$ is an \typeL{} suffix.
The substring between two succeeding $\typeHoshi$ suffixes is called an \teigi{LMS substring}. 
In other words, a substring $T[i..j]$ with $i < j$ is an LMS substring if and only if $T[i..|T|]$ and $T[j..|T|]$ are \typeHoshi{} suffixes and 
there is no $k \in (i..j)$ such that $T[k..|T|]$ is an \typeHoshi{} suffix.
A border case is $T[|T|..|T|]$, which has to be the smallest suffix of $T$ (and can be achieved by appending the artificial character~$\texttt{\$}$ to $T$ lexicographically smaller than all other characters appearing it~$T$) such that $T||T|..|T|]$ in an \typeHoshi{} suffix.
We additionally treat $T[|T|..|T|]$ as an LMS substring.
The types for the suffixes of our running example are given in \cref{figExLMSSAIS}.
Regarding the defined types, we make no distinction between suffixes and their starting positions
(e.g., the statements that (a) $T[i]$ is type~\typeL{} and (b) $T[i..|T|]$ is an \typeL{} suffix are equivalent).

Next, {Nong et al.}~\cite[Def.~3.3]{nong11sais} define a relation~\teigi{$\LMSOrder$} on substrings of $T$ based on the lexicographic order and the types:
Given two substrings~$S$ and~$U$.
Let $i$ be the smallest integer such that 
(1) $S[i] < U[i]$ or 
(2) $S[i]$ is type $\typeL$ and $U[i]$ is type~$\typeS$ or~$\typeHoshi$.
If such an $i$ exists, then we write $S \LMSOrder U$.
For two LMS substrings $S$ and~$U$ with $S \not= U$, either $S \LMSOrder U$ or $U \LMSOrder S$,
even if $S$ is a prefix of $U$ (cf.\ the discussion below of Def.~3.3 in~\cite{nong11sais}).
So $\LMSOrder$ is an order on the LMS substrings.
The $\LMSOrder$-order is shown on the left side of \cref{figExRankingSAIS} for the LMS substrings listed of the left side of \cref{figExLMSSAIS}.
The crucial observation is that the $\LMSOrder$-order of the LMS substrings coincides with the lexicographic order of the suffixes starting with the LMS substrings~\cite[Lemma 3.8]{nong11sais}.

\begin{figure}
    \centering{\begin{tabular}{lll}
		    \toprule
		    LMS Substring & Contents & Non-Terminal
		    \\\midrule
			{$T[2..5]$}    & \texttt{bbca}  & \texttt{E} \\
			{$T[5..7]$}         & \texttt{acb}  & \texttt{C} \\
			{$T[7..10]$}    & \texttt{bbca}  & \texttt{E} \\
			{$T[10..12]$}        & \texttt{ada} & \texttt{D} \\
			{$T[12..15]$}  & \texttt{acba}  & \texttt{B} \\
			{$T[15..17]$}       & \texttt{ada}  & \texttt{D} \\
			{$T[17..20]$} & \texttt{acba} & \texttt{B} \\
			{$T[20..20]$} & \texttt{a} & \texttt{A} \\
		    \bottomrule
		\end{tabular}
		\hfill
	\begin{tabular}{ll}
	    \toprule
	    \typeHoshi{} Suffix & Contents
	    \\\midrule
	    $T[20]$      & $\texttt{a}$ \\
	    $T[17..20] $ & $\texttt{acba}$ \\
	    $T[12..20] $ & $\texttt{acbadacba}$ \\
	    $T[5..20]$   & $\texttt{acbbcadacbadacba}$ \\
	    $T[15..20] $ & $\texttt{adacba}$ \\
	    $T[10..20] $ & $\texttt{adacbadacba}$ \\
	    $T[2..20]$   & $\texttt{bbcacbbcadacbadacba}$ \\
	    $T[7..20] $  & $\texttt{bbcadacbadacba}$ \\
	    \bottomrule
	\end{tabular}
    }\caption{Ranking of the LMS substrings and the \typeHoshi{} suffixes of our running example given in \cref{secExample} and \cref{figExLMSSAIS}.
      \emph{Left}: LMS substrings assigned with non-terminals reflecting their corresponding rank in \LMSOrder{}-order.
      \emph{Right}: \typeHoshi{} suffixes of $T$ sorted in \LexOrder{}-order.
    Note that $T[5..7] = \texttt{acb} \LexOrder \texttt{acba} = T[12..15] = T[17..20]$, but $\texttt{acba} \LMSOrder \texttt{acb}$.
}
\label{figExRankingSAIS}
\end{figure}

	\begin{figure*}
	  \centering{\begin{tikzpicture}

  \matrix[InductionMatrixSubstring] (sa) {1      & 2 & 3  & 4  & 5  & 6  & 7 & 8 & 9 & 10 & 11 & 12 & 13 & 14 & 15 & 16 & 17 & 18 & 19 & 20 & 3\\
  20     & 5 & 10 & 12 & 15 & 17 &   &   & 2 & 7  &    &    &    &    &    &    &    &    &    &    & 4\\
         &   &    &    &    &    & 19&14 &   &    &    &    & 4  & 9  & 18 & 13 & 1  &  6 & 11 & 16 & 5\\
	 & &  &   &  &  &   &   &  &   &   &   &    &    &    &    &    &    &    &    & \\
	 &17 & 12 & 5  & 10 & 15 &   &   & 2 & 7  & 3  &  8 &    &    &    &    &    &    &    &    & 6\\
	 & &  &   &  &  &   &   &  &   &   &   &    &    &    &    &    &    &    &    & \\
         & \texttt{A} &  \texttt{A} & \texttt{B}  &  \texttt{C} & \texttt{C}  &   &   & \texttt{D} & \texttt{D}  &    &    &    &    &    &    &    &    &    &    & 7 \\
	};
	\node [left=0cm of sa-2-1] {\typeHoshi{} suffixes};
	\node [left=0cm of sa-3-1] {\typeL{} suffixes};
	\node [left=0cm of sa-5-1] {\typeS{} suffixes};
	\node [left=0cm of sa-7-1] {$\LMSOrder$-ranks};
\setcounter{PfeilCount}{1}
	\Pfeil{2}{1}{3}{7}{0.2em}
	\Pfeil{2}{2}{3}{13}{0.9em}
	\Pfeil{2}{3}{3}{14}{0.5em}
	\Pfeil{2}{4}{3}{19}{1.0em}
	\Pfeil{2}{5}{3}{8}{0.7em}
	\Pfeil{2}{6}{3}{20}{1.1em}

	\Pfeil{3}{7}{3}{15}{1.2em}
	\Pfeil{3}{8}{3}{16}{1.5em}
	\Pfeil{2}{9}{3}{17}{1.7em}
	\Pfeil{2}{10}{3}{18}{2em}
        
	\Pfeil{4}{20}{5}{6}{0.8em}
	\Pfeil{4}{19}{5}{5}{0.2em}
	\Pfeil{4}{18}{5}{4}{0.3em}
	\Pfeil{4}{16}{5}{3}{0.4em}
	\Pfeil{4}{15}{5}{2}{0.5em}
	\Pfeil{4}{14}{5}{12}{0.6em}
	\Pfeil{4}{13}{5}{11}{0.7em}

	\Pfeil{5}{12}{5}{10}{1.5em}
	\Pfeil{5}{11}{5}{9}{1.3em}

	\tikzstyle{Schrift} = [minimum height=1em, text height=0.5em, text depth=0.1pt, minimum width=2em, draw]

	\node [yshift=0.5em, fit=(sa-1-1.north) (sa-1-6.north), Schrift] (aS) {\typeS};
	\node [yshift=0.5em, fit=(sa-1-7.north) (sa-1-8.north), Schrift] (bL) {\typeL};
	\node [yshift=0.5em, fit=(sa-1-9.north) (sa-1-12.north),Schrift] (bS) {\typeS};
	\node [yshift=0.5em, fit=(sa-1-13.north) (sa-1-18.north),Schrift] (cL) {\typeL};
\node [yshift=0.5em, fit=(sa-1-19.north) (sa-1-20.north),Schrift] (dL) {\typeL};

	\node [yshift=2em, fit=(sa-1-1.north) (sa-1-6.north), Schrift] (a) {\texttt{a}};
	\node [yshift=2em, fit=(sa-1-7.north) (sa-1-12.north), Schrift] {\texttt{b}};
	\node [yshift=2em, fit=(sa-1-13.north) (sa-1-18.north), Schrift] {\texttt{c}};
	\node [yshift=2em, fit=(sa-1-19.north) (sa-1-20.north), Schrift] {\texttt{d}};
	\node [above=0em of sa-1-21] {\tiny{2}};
	\node [above=1.5em of sa-1-21] {\tiny{1}};

	\node [left=0cm of aS] {types};
	\node [left=0cm of a,align=left,yshift=1em] {starting\\character};

\end{tikzpicture}
	  }\caption{Inducing LMS substrings.
	    Rows~1 and~2 show the partitioning of $\SA$ into buckets, first divided by the starting characters of the respective LMS substrings, and second by the types \typeL{} and \typeS{}. 
	    In Row~4, the \typeHoshi{} suffixes are inserted into their respective \typeS{} buckets. 
	    Here it is sufficient to only put the smallest \typeHoshi{} suffix in the correct order among all other \typeHoshi{} suffixes in the same bucket. This suffix is $T[20..20]$ in our example, stored at the suffix array entry~$1$.
	The \typeHoshi{} (resp.\ \typeL{}) suffixes induce the \typeL{} (resp.\ \typeS{}) suffixes in Row~5 (resp.\ Row~6).
	    The last row assigns each \typeHoshi{} suffix a meta-character representing its \LMSOrder{}-rank.
	    We can compute two subsequent suffixes by character-wise comparison, spending \Oh{|T|} time in total since the LMS substrings have a total length of \Oh{|T|}.
	  }
	  \label{figInducingLMS}
	\end{figure*}
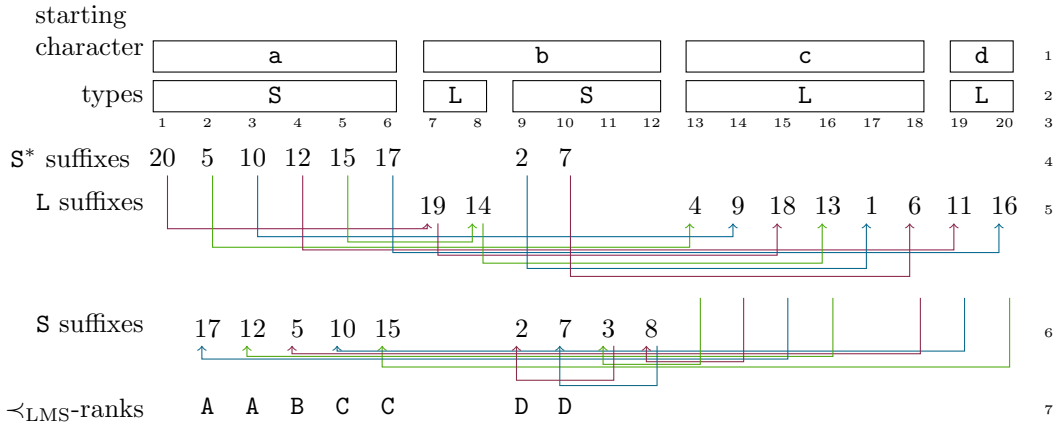

	\begin{figure*}
	  \centering{\begin{tikzpicture}

\matrix[InductionMatrix] (sa) {1    & 2  & 3  & 4 & 5  & 6  & 7  & 8  & 9 & 10 & 11 & 12 & 13 & 14 & 15 & 16 & 17 & 18 & 19 & 20 & 3\\
  20   & 17 & 12 & 5 & 15 & 10 &    &    & 2 & 7  &    &    &    &    &    &    &    &    &    &    & 4\\
       &    &    &   &    &    & 19 & 14 &   &    &    &    & 4  & 9  & 18 & 13 & 1  & 6  & 16 & 11 & 5\\
       &    &    &   &    &    &    &    &   &    & 3  & 8  &    &    &    &    &    &    &    &    & 6 \\
       &    &    &   &    &    &    &    &   &    &    &    &    &    &    &    &    &    &    &    & \\
  20   & 17 & 12 & 5 & 15 & 10 & 19 & 14 & 2 & 7  & 3  & 8  & 4  & 9  & 18 & 13 & 1  & 6  & 16 & 11 & 7\\
  19   & 16 & 11 & 4 & 14 & 9  & 18 & 13 & 1 & 6  & 2  & 7  & 3  & 8  & 17 & 12 & 20 & 5  & 15 & 10  & 8\\
  b    & d  & d  & c & b  & c  & c  & c  & c & c  & b  & b  & b  & b  & a  & a  & a  & a  & a  & a  & {\tiny 9}\\
	};
	\node [left=0cm of sa-2-1] {\typeHoshi{} suffixes};
	\node [left=0cm of sa-3-1] {\typeL{} suffixes};
	\node [left=0cm of sa-4-1] {\typeS{} suffixes};
	\node [left=0cm of sa-8-1] {$\BWT = $};
	\node [left=0cm of sa-7-1] {$\SA-1= $};
	\node [left=0cm of sa-6-1] {$\SA= $};

	\setcounter{PfeilCount}{1}
	\Pfeil{2}{1}{3}{7}{0.2em}
	\Pfeil{2}{2}{3}{19}{3em}
	\Pfeil{2}{3}{3}{20}{2.8em}
	\Pfeil{2}{4}{3}{13}{2em}
	\Pfeil{2}{5}{3}{8}{4em}
	\Pfeil{2}{6}{3}{14}{2.4em}

	\Pfeil{2}{9}{3}{17}{5.2em}
	\Pfeil{2}{10}{3}{18}{5.6em}
\Pfeil{3}{7}{3}{15}{4.5em}
	\Pfeil{3}{8}{3}{16}{4em}
\Pfeil{3}{14}{4}{12}{4em}
	\Pfeil{3}{13}{4}{11}{2em}

	\tikzstyle{Schrift} = [minimum height=1em, text height=0.5em, text depth=0.1pt, minimum width=2em, draw]

	\node [yshift=0.5em, fit=(sa-1-1.north) (sa-1-6.north), Schrift] (aS) {\typeS};
	\node [yshift=0.5em, fit=(sa-1-7.north) (sa-1-8.north), Schrift] (bL) {\typeL};
	\node [yshift=0.5em, fit=(sa-1-9.north) (sa-1-12.north),Schrift] (bS) {\typeS};
	\node [yshift=0.5em, fit=(sa-1-13.north) (sa-1-18.north),Schrift] (cL) {\typeL};
\node [yshift=0.5em, fit=(sa-1-19.north) (sa-1-20.north),Schrift] (dL) {\typeL};

	\node [yshift=2em, fit=(sa-1-1.north) (sa-1-6.north), Schrift] (a) {\texttt{a}};
	\node [yshift=2em, fit=(sa-1-7.north) (sa-1-12.north), Schrift] {\texttt{b}};
	\node [yshift=2em, fit=(sa-1-13.north) (sa-1-18.north), Schrift] {\texttt{c}};
	\node [yshift=2em, fit=(sa-1-19.north) (sa-1-20.north), Schrift] {\texttt{d}};
	\node [above=0em of sa-1-21] {\tiny{2}};
	\node [above=1.5em of sa-1-21] {\tiny{1}};

	\node [left=0cm of aS] {types};
	\node [left=0cm of a,align=left,yshift=1em] {starting\\character};

\end{tikzpicture}
	  }\caption{Inducing \typeL{} and \typeS{} suffixes from the \LexOrder-order of the \typeHoshi{} suffixes given in \cref{figExLMSSAIS}.
	    Rows~1 and~2 show the partitioning of $\SA$ into buckets, first divided by the starting characters of the respective suffixes, and second by the types \typeL{} and \typeS{}. Row~4 is $\SA$ after inserting the \typeHoshi{} suffixes according to their \LexOrder{}-order rank obtained from the right of \cref{figExRankingSAIS}. 
	The \typeHoshi{} (resp.\ \typeL{}) suffixes induce the \typeL{} (resp.\ \typeS{}) suffixes in Row~5 (resp.\ Row~6).
	Putting all together yields \SA{} in Row~7. 
	In the penultimate row $\SA{}-1$, each text position stored in \SA{} is decremented by one, or set to~$n$ if this position was~$1$.
	The last row shows $T[(\SA - 1)[i]] = \BWT[i]$ in its $i$-th column, which is the BWT of~$T$.
	This BWT is not reversible since the input is not terminated with a unique character like~$\texttt{\$}$.
	To obtain the BWT of~$T\texttt{\$}$, we first write $T[\SA[1]] = T[20] = \texttt{a}$ to the output, and then \BWT, but exchanging $\BWT[\SA^{-1}[1]] = \BWT[17] = \texttt{a}$  with \texttt{\$},
	i.e., \texttt{abddcbcccccbbbbaa\$aaa}.
      }
	  \label{figExInduceSAIS}
	\end{figure*}

{Nong et al.}~\cite[A3.4]{nong11sais} compute the $\LMSOrder$-order of all LMS substrings with the induced sorting (which we describe below for the step of computing the rank of all suffixes).
\Cref{figInducingLMS} visualizes this computation on our running example.
Hence, we can assign each LMS substring a rank based on the $\LMSOrder$-order.
Next, we build a string~$T^{(1)}$ of LMS substring ranks with $T^{(1)}[i]$ being the rank of the $i$-th LMS substring of~$T$ in text order.\footnote{We can obtain $T^{(1)}$ by 
scanning $T$ from left to right and 
replacing each LMS substring by its respective rank, but keep its last character in $T$ if this character is the first character of the subsequent LMS substring.
We further omit the first characters of~$T$ that are not part of an LMS substring (which must be of type~\typeL{}).}
See the right side of \cref{figExLMSSAIS} for our running example.
We recursively call SAIS on this text of ranks 
until the ranks of all LMS substrings are distinct.
Given that we have computed $T^{(k)}$ and all characters of $T^{(k)}$ (i.e., the ranks of the respective LMS substrings) are distinct,
then these ranks determine the order of the \typeHoshi{} suffixes of~$T^{(k)}$.
The order of the \typeHoshi{} suffixes of our running example are given in \cref{figExRankingSAIS} on the right side.
Having the order of the \typeHoshi{} suffixes, 
we allocate space for the suffix array, and divide the suffix array into buckets, 
grouping each suffix with the same starting character and same type (either \typeL{} or \typeS{}) into one bucket.
Among all suffixes with the same starting character, the \typeL{} suffixes precede the \typeS{} suffixes~\cite[Corollary~3]{ko05sa}.
Putting \typeHoshi{} suffixes in their respective buckets according to their order (smallest elements are the leftmost elements in the buckets),
we can induce the \typeL{} suffixes, as these precede either \typeL{} or \typeHoshi{} suffixes.
For that, we scan $\SA{}$ from left to right, and take action only for suffix array entries that are not empty:
When accessing the entry $\SA[k] = i$ with $i > 1$, write $i-1$ to the leftmost available slot of the \typeL{} bucket with the character $T[i-1]$ if $T[i-1..|T|]$ is an \typeL{} suffix.
Finally, we can induce the $\LexOrder$-order of the \typeS{} suffixes by scanning the suffix array from right to left:
When accessing the entry $\SA[k] = i$, write $i-1$ to the rightmost available slot of the \typeS{} type bucket with the character $T[i-1]$ if $T[i-1..|T|]$ is an \typeS{} suffix.
As an invariant, we always fill an \typeL{} bucket and an \typeS{} bucket from left to right and from right to left, respectively.
So we can think of each \typeL{} bucket and each \typeS{} bucket as a list with an insertion operation at the end or at the beginning, respectively.
We conduct these steps for our running example in \cref{figExInduceSAIS}.

\begin{figure}
  \centerline{\begin{tabular}{lllll}
      \toprule
      $U$ & $V$    & \LexOrder{} & \OmegaOrder{} & \LMSOrder{}
      \\\midrule
      \texttt{aba} & \texttt{aca}    & $<$ & $<$ & $<$ \\
      \texttt{adc} & \texttt{adcb}   & $<$ & $<$ & $>$ \\
      \texttt{acb} & \texttt{acba}   & $<$ & $>$ & $>$ \\
      \bottomrule
    \end{tabular}
  }\caption{Comparison of the three orders studied in this paper applied to LMS substrings.
  Assume that $U$ and $V$ are substrings of the text surrounded by a character~$\texttt{d}$ 
  (i.e., $T = \ldots \texttt{d}U\texttt{d} \ldots \texttt{d}V\texttt{d} \ldots$) 
  such that the first and the last character of both~$U$ and~$V$ start with an \typeHoshi{} suffix. We mark with the signs $<$ and $>$ whether $U$ is smaller or respectively larger than $V$ according to the corresponding order. The orders can differ only when one string is the prefix of another string, as this is the case in the last two rows.
  Finally, occurrences of $U$ and $V$ can be $\LMSOrder$-incomparable
  in different contexts such as 
  $\ldots \texttt{d}U\texttt{a} \ldots \texttt{d}V\texttt{d} \ldots$, for instance.
}
  \label{tableOrderDiffer}
\end{figure}

In total, the induction takes \Oh{|T|} time.
The recursion step takes also \Oh{|T|} time since there are at most $|T|/2$ LMS substrings 
(there are no two text positions~$T[i]$ and $T[i+1]$ with type~\typeHoshi{} for $i \in [1..n-1]$).
This gives $\mathcal{T}(n) = \mathcal{T}(n/2) + \Oh{n} = \Oh{n}$ total time,
where $\mathcal{T}(n)$ denotes the time complexity for computing a suffix array of length~$n$.

However, with SAIS we cannot obtain $\CSA$ ad-hoc since we need to exchange \LexOrder{} with \OmegaOrder{}.
Although these orders are the same for Lyndon words~\cite[Thm.~8]{bonomo14sorting}, they differ for LMS substrings as can be seen in \cref{tableOrderDiffer}.
Hence, we need to come up with an idea to modify SAIS in such way to compute \CSA{}.

\FloatBarrier
\section{Our Adaptation}
We want SAIS to sort Lyndon conjugates in \OmegaOrder{}-order instead of suffixes in \LexOrder{}-order.
For that, we first get rid of duplicate Lyndon factors to facilitate the analysis, and then subsequently
introduce a slightly different notion to the types of suffixes and LMS substrings,
which translates the suffix sorting problem into computing the BBWT\@.

\subsection{Reduced String and Composed Lyndon Factorization}\label{secComposedString}
In a pre-computation step, we want to facilitate our analysis by removing all identical Lyndon factors from~$T$ yielding a reduced string~$R$.
We want to remove them to make conjugates unique; thus we can linearly order them.
Consequently, the first step is to show that we can obtain the BBWT of~$T$ from the circular suffix array of~$R$ (which we will subsequently define):

The (composed) \teigi{Lyndon factorization}~\cite{chen58lyndon} of $T \in\Sigma^+$ is 
the factorization of $T$ into $T_1^{\tau_1} \cdots T_t^{\tau_t} = T$,
where $T_1, \ldots, T_t$ is a sequence of lexicographically decreasing Lyndon words and $\tau_x \ge 1$ for $x \in [1..t]$.
Let $R := T_1 \cdots T_t$ denote the text, in which all duplicate Lyndon factors are removed.
Obviously, the Lyndon factorization of $R$ is $T_1, \ldots, T_t$.
Let $\ibeg{T_x}$ and $\iend{T_x}$ denote the starting and ending position of the $x$-th Lyndon factor in $R$, i.e.,
$R[\ibeg{T_x}..\iend{T_x}]$ is the $x$-th Lyndon factor~$T_x$ of~$R$.

\newcommand*{\ConjSet}{\ensuremath{\mathcal{S}}}
Our aim is to compute the $\OmegaOrder$-order of all conjugates of all Lyndon factors of~$R$, which are given by the set 
$\ConjSet := \bigcup_{x = 1}^{t} \conj{T_x}$.
Like {Hon et al.}~\cite{hon12ebwt}, we present this order in the so-called \teigi{circular suffix array} $\CSA$ of $\{T_1, \ldots, T_t\}$, i.e.,
an array of length $|R|$ with $\CSA[k] = i$ if $R[i..\iend{T_x}]R[\ibeg{T_x}..i-1]$ is the $k$-th smallest
string in $\ConjSet$ with respect to $\OmegaOrder$, where $i \in [\ibeg{T_x}..\iend{T_x}]$.
The length of $\CSA$ is $|R|$ since we can associate each text position~$\CSA[k]$ in~$R$ with a conjugate starting with~$R[\CSA[k]]$.

Having the circular suffix array~$\CSA$ of $\{T_1, \ldots, T_t\}$, we can compute the BBWT of $T$ 
by reading $\CSA[k]$ for $k \in [1..|R|]$ from left to right:
Given $\CSA[k] = i \in [\ibeg{T_x}..\iend{T_x}]$, 
we append $T[i^-]$ exactly $\tau_x$ times to $\BBWT$, 
where $i^-$ is $i - 1$ or $\iend{T_x}$ if $i = \ibeg{T_x}$.
(This is analogous to the definition of BWT where we set $\BWT[i] = T[n]$ for $\SA[i] = 1$, but here we wrap around each Lyndon factor.)

\subsection{Translating Types to Inf-Suffixes}
In what follows, we continue working with $R$ defined in \cref{secComposedString} instead of $T$.
Let $R[i..]$ denote the infinite string $R[i..\iend{T_x}]T_xT_x \cdots = \conj[k]{T_x} \conj[k]{T_x} \cdots$ with $x$ such that $i \in [\ibeg{T_x}..\iend{T_x}]$
and $k = i - \ibeg{T_x}$.
We say that $R[i..]$ is an \teigi{inf-suffix}.
As a shorthand, we also write $T_x[i..] = \conj[i-1]{T_x}  \conj[i-1]{T_x} \cdots$ for the inf-suffix starting at $R[\ibeg{T_x}+i-1]$.
In particular, $T_x[|T_x|+1..] = T_x[1..] = T_x T_x \cdots$.

Like in SAIS, we distinguish between $\typeL$ and $\typeS$ inf-suffixes:
\begin{itemize}
	\item $R[i..]$ is an $\typeL$ inf-suffix if $R[i..] \LexOrderSucc R[i^+..]$, and
	\item $R[i..]$ is an $\typeS$ inf-suffix if $R[i..] \LexOrder R[i^+..]$,
\end{itemize}
where $i^+$ is either $i+1$ or $\ibeg{T_x}$ if $i = \iend{T_x}$,
and~$x$ is given such that $i \in [\ibeg{T_x}..\iend{T_x}]$.
Finally, we introduce the $\typeHoshi$ inf-suffixes as a counterpart to the $\typeHoshi$ suffixes:
If $R[i..]$ is an $\typeS$ inf-suffix, it is further an $\typeHoshi$ inf-suffix if 
$R[i^-..]$ is an $\typeL$ inf-suffix with $i^-$ being either $i-1$ or \iend{T_x} if $i = \ibeg{T_x}$, and $x \in [1..t]$ chosen such that $i \in [\ibeg{T_x}..\iend{T_x}]$.

When speaking about types, we do not distinguish between an inf-suffix and its starting position in $R$.
This definition assigns all positions of~$R$ a type except those belonging to a Lyndon factor of length one.
We solve this by stipulating that all Lyndon factors of length one start with an \typeHoshi{} inf-suffix.
However, in what follows, we temporarily omit all Lyndon factors of length one because we will later
see that they can be placed at the beginning of their corresponding buckets in the circular suffix array.
They nevertheless appear in the examples for completeness.
To show that suffixes and inf-suffixes starting at the same position have the same type (except for some border-cases),
the following lemma will be particularly useful:
\begin{lemma}[{\cite[Lemma~7]{bonomo14sorting}}]\label{lemLyndonOrderEquiv}
    For $i,j \in [1..|T_x|]$ and $x \in [1..t]$, the following statements are equivalent:
    \begin{enumerate}
    \item $\conj[i-1]{T_x} = T_x[i..|T_x|]T_x[1..i-1] \LexOrder T_x[j..|T_x|]T_x[1..j-1] = \conj[j-1]{T_x}$;
    \item $\conj[i-1]{T_x} \OmegaOrder \conj[j-1]{T_x}$, i.e., $T_x[i..] \LexOrder T_x[j..]$;
      \item $T_x[i..|T_x|] \LexOrder T_x[j..|T_x|]$.
    \end{enumerate}
\end{lemma}

\begin{lemma}
  Omitting all Lyndon factors of length one from~$R$, the types of all positions match the original SAIS types, 
  except maybe $R[1]$ and $R[\ibeg{T_t}+1..|R|]$, 
  where $R[1..]$ and $R[|R|..|R|]$ are always an \typeHoshi{} inf-suffix and an \typeHoshi{} suffix, respectively.
\end{lemma}
\begin{proof}
  We show that inf-suffixes as well as suffixes starting with Lyndon factors have the same type $\typeHoshi$:
\begin{description}
  \item[inf-suffxes.]
    Assume that $R[\ibeg{T_x}..]$ is an \typeL{} inf-suffix for an $x \in [1..t]$.
    According to the definition $R[\ibeg{T_x}+1..] \LexOrder R[\ibeg{T_x}..]$, i.e., $T_x[2..] \LexOrder T_x[1..]$,
    and with \cref{lemLyndonOrderEquiv}, $T_x[2..|T_x|] \LexOrder T_x$, contradicting that $T_x$ is a Lyndon word.
    Finally, $R[\ibeg{T_x}..]$ is an \typeHoshi{} inf-suffix because $T_x \LexOrder T_x[|T_x|]$ and hence $T_x[1..] \LexOrder T_x[|T_x|..]$, again with \cref{lemLyndonOrderEquiv}.
  \item[suffixes.] 
Due to the Lyndon factorization, $R[\ibeg{T_x}..|R|] \LexOrderSucc R[\ibeg{T_{x+1}}..|R|]$ for $x \in [1..t-1]$.
Hence, the suffix $R[\iend{T_x}..|R|]$ starting at $R[\iend{T_x}]$ has to be lexicographically larger than the suffix $R[\iend{T_x}+1..|R|] = R[\ibeg{T_{x+1}}..|R|]$, 
  otherwise we could extend the Lyndon factor~$T_x$.

\end{description}
Consequently, $R[\ibeg{T_x}..|R|]$ and $R[\ibeg{T_x}..]$ are an $\typeHoshi$ suffix and an $\typeHoshi$ inf-suffix, respectively,
and $R[\iend{T_x}..|R|]$ and $R[\iend{T_x}..]$ are an $\typeL$ suffix and an $\typeL$ inf-suffix.

  The claim for all other positions ($\bigcup_{x=1}^{t-1}[\ibeg{T_x}+1..\iend{T_x}-1]$) follows by 
  observing that $T_x[1..]$ is the $\LexOrder$-smallest inf-suffix among all inf-suffixes starting in $T_x$ and
  $R[\ibeg{T_{x+1}}..|R|]$ is $\LexOrder$-smaller than all suffixes starting in $R[\ibeg{T_x}..\iend{T_x}]$ for $x \in [1..t-1]$.
\end{proof}
A corollary is that $R[i..|R|] \LexOrder R[i..]$ for $i \in [\ibeg{T_x}..\iend{T_x}]$ and $x \in [1..t-1]$ since $T_{x+1} \LexOrder T_x$.\footnote{Consequently, for transforming \SA{} into \CSA{}, one only needs to shift values in \SA{} to the right, as this is done by one of the implementations mentioned in the related work.}
Next, we define the equivalent to the LMS substrings for the inf-suffixes, which we call \emph{LMS inf-suffixes}:
For $1 \le i < j \le |T_x|+1$, the substring $(T_xT_x)[i..j]$ is called an \teigi{LMS inf-substring} if and only if $T_x[i..]$ and $T_x[j..]$ are $\typeHoshi$ inf-suffixes and
there is no $k \in (i..j)$ such that $T_x[k..]$ is an \typeHoshi{} inf-suffix.
This definition differs from the original LMS substrings (omitting the last one $R[|R|..|R|]$ being a border case)
only for the last LMS inf-substring of each Lyndon factor.
Here, we append $T_x[1]$ instead of $T_{x+1}[1]$ to the suffix starting with the last type~\typeHoshi{} position of $T_x$.

\subsection{Example}\label{secExample}

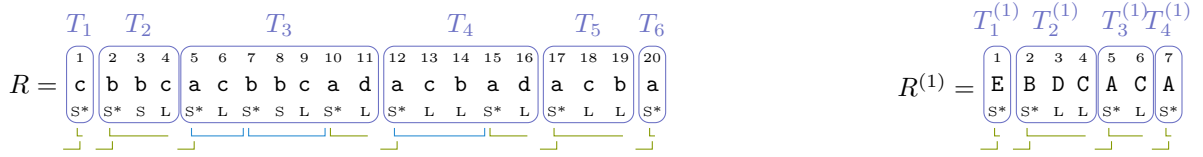
\begin{figure}
    \centering{\newcommand*{\LMSleftbracket}[2]{\begin{scope}[transform canvas={yshift=-0.5em}]
		\draw [color=solarizedGreen] ([yshift=0.3em,xshift=-0.1em]text-3-#1.south) -- ([xshift=-0.1em]text-3-#1.south) -- ([xshift=0.1em]text-3-#2.south) {};
	\end{scope}
}
	\newcommand*{\LMSrightbracket}[1]{\begin{scope}[transform canvas={yshift=-1em}]
		\draw [color=solarizedGreen] (text-3-#1.south west) -- (text-3-#1.south) -- ([yshift=0.3em]text-3-#1.south) {};
	\end{scope}
}

\begin{tikzpicture}
\matrix[LMSmatrix] (text) {1 & 2  & 3 & 4 & 5  & 6 & 7  & 8 & 9 & 10 & 11 & 12 & 13 & 14 & 15 & 16 & 17 & 18 & 19 & 20 \\
c  & b  & b & c & a  & c & b  & b & c & a  & d  & a  & c  & b  & a  & d  & a  & c  & b  & a \\
S* & S* & S & L & S* & L & S* & S & L & S* & L  & S* & L  & L  & S* & L  & S* & L  & L  & S*  \\
	};
	\node [fit=(text-1-1.north) (text-3-1.south), LyndonBlock, color=solarizedViolet, label={[color=solarizedViolet]above:$T_{1}$}] {};
	\node [fit=(text-1-2.north) (text-3-4.south), LyndonBlock, color=solarizedViolet, label={[color=solarizedViolet]above:$T_{2}$}] {};
	\node [fit=(text-1-5.north) (text-3-11.south), LyndonBlock, color=solarizedViolet, label={[color=solarizedViolet]above:$T_{3}$}] {};
	\node [fit=(text-1-12.north) (text-3-16.south), LyndonBlock, color=solarizedViolet, label={[color=solarizedViolet]above:$T_{4}$}] {};
	\node [fit=(text-1-17.north) (text-3-19.south), LyndonBlock, color=solarizedViolet, label={[color=solarizedViolet]above:$T_{5}$}] {};
	\node [fit=(text-1-20.north) (text-3-20.south), LyndonBlock, color=solarizedViolet, label={[color=solarizedViolet]above:$T_{6}$}] {};
	\node [left=0cm of text-2-1] {$R = $};

	\LMSleftbracket{1}{1}
	\LMSrightbracket{1}

	\LMSleftbracket{2}{4}
	\LMSrightbracket{2}

	\LMSleftbracket{10}{11}
	\LMSrightbracket{5}

	\LMSleftbracket{15}{16}
	\LMSrightbracket{12}

	\LMSleftbracket{17}{19}
	\LMSrightbracket{17}

	\LMSleftbracket{20}{20}
	\LMSrightbracket{20}
	
\LMSbracket{5}{7}
\LMSbracket{7}{10}
\LMSbracket{12}{15}

\end{tikzpicture}
\hfill
\begin{tikzpicture}
\matrix[LMSmatrix] (text) {1 & 2 & 3 & 4 & 5 & 6 & 7 \\
    E & B & D & C & A & C & A \\
    S* & S* & L & L & S*  & L & S* \\
	};
	\node [fit=(text-1-1.north) (text-3-1.south), LyndonBlock, color=solarizedViolet, label={[color=solarizedViolet]above:$T^{(1)}_{1}$}] {};
	\node [fit=(text-1-2.north) (text-3-4.south), LyndonBlock, color=solarizedViolet, label={[color=solarizedViolet]above:$T^{(1)}_{2}$}] {};
	\node [fit=(text-1-5.north) (text-3-6.south), LyndonBlock, color=solarizedViolet, label={[color=solarizedViolet]above:$T^{(1)}_{3}$}] {};
	\node [fit=(text-1-7.north) (text-3-7.south), LyndonBlock, color=solarizedViolet, label={[color=solarizedViolet]above:$T^{(1)}_{4}$}] {};
	\node [left=0cm of text-2-1] {$R^{(1)} = $};

	\LMSleftbracket{1}{1}
	\LMSrightbracket{1}

	\LMSleftbracket{2}{4}
	\LMSrightbracket{2}

	\LMSleftbracket{5}{6}
	\LMSrightbracket{5}

	\LMSleftbracket{7}{7}
	\LMSrightbracket{7}
\end{tikzpicture}
\vspace{1em}
    }\caption{Splitting $R$ and $R^{(1)}$ into LMS inf-substrings. 
	The rectangular brackets below the types represent the LMS inf-substrings.
	Broken brackets denote that the corresponding LMS inf-substring ends with the first character of the Lyndon factor in which it is contained.
	They are colored in green~\protect\PatternLegend{fill=solarizedGreen};
	all other LMS inf-substrings are represented by brackets colored in blue~\protect\PatternLegend{fill=solarizedBlue}.
	$R^{(1)}$ is $R$ after the replacement of its LMS inf-substrings with their corresponding ranks defined in \cref{secExample} and on the left of \cref{figExRanking}.
    }
    \label{figExLMS}
\end{figure}

\begin{figure}
    \centering{\begin{tabular}{lll}
		    \toprule
		    LMS Inf-Substring & Contents & Non-Terminal
		    \\\midrule
		{\color{solarizedGreen}$R[1]R[1]$}    & \texttt{cc}  & - \\
		{\color{solarizedGreen}$R[2..4]R[2]$}    & \texttt{bbcb}  & \texttt{E} \\
		{\color{solarizedBlue}$R[5..7]$}         & \texttt{acb}  & \texttt{B} \\
		{\color{solarizedBlue}$R[7..10]$}         & \texttt{bbca}  & \texttt{D} \\
		{\color{solarizedGreen}$R[10..11]R[10]$}    & \texttt{ada}  & \texttt{C} \\
		{\color{solarizedBlue}$R[12..15]$}        & \texttt{acba} & \texttt{A} \\
		{\color{solarizedGreen}$R[15..16]R[12]$}  & \texttt{ada}  & \texttt{C} \\
		{\color{solarizedGreen}$R[17..19]R[17]$} & \texttt{acba} & \texttt{A} \\
		{\color{solarizedGreen} $R[20]R[20]$} &  \texttt{aa} & - \\
		    \bottomrule
		\end{tabular}
		\hfill
	\begin{tabular}{ll}
	    \toprule
	    \typeHoshi{} Inf-Suffix & Contents
	    \\\midrule
	    $R[20..]$  & $\texttt{a}\ldots$ \\
	    $R[17..]$  & $\texttt{acb}\ldots$ \\
	    $R[12..] $  & $\texttt{acbad}\ldots$ \\
	    $R[5..] $  & $\texttt{acbbcad}\ldots$ \\
	    $R[15..] $   & $\texttt{adacb}\ldots$ \\
	    $R[10..] $  & $\texttt{adacbbc}\ldots$ \\
	    $R[7..] $   & $\texttt{bbcadac}\ldots$ \\
	    $R[2..] $ & $\texttt{bbc}\ldots$ \\
	    $R[1..]  $  & $\texttt{c}\ldots$ \\
	    \bottomrule
	\end{tabular}
    }\caption{Ranking of the LMS inf-substrings and the \typeHoshi{} suffixes of our running example $T = R$ given in \cref{secExample} and \cref{figExLMS}.
      \emph{Left}: LMS inf-substrings assigned with non-terminals reflecting their corresponding rank in \LMSOrder{}-order.
      They have the same color as the respective rectangular brackets on the left of \cref{figExLMS}.
      The first and the last LMS substring do not receive a non-terminal since their lengths are one
      (remember that we omit Lyndon factors of length~$1$ in the recursive call).
      \emph{Right}: \typeHoshi{} inf-suffixes of $T$ sorted in \LexOrder{}-order, which corresponds to the \OmegaOrder{} of the conjugate starting with this inf-suffix.
    Compared with \cref{figExRankingSAIS}, the suffixes $R[2..20]$ and $R[7..20]$ in the \LexOrder{}-order are order differently than their respective inf-suffixes $R[2..]$ and $R[7..]$ in the \LexOrder{}-order.
}
    \label{figExRanking}
\end{figure}

The LMS inf-substrings of our running example $T := \texttt{\RunningExample{}}$ with $R = T$ are given in \cref{figExLMS}.
Their \LMSOrder{}-ranking is given on the left side of \cref{figExRanking}, 
	where we associate each LMS inf-substring, except those consisting of a single character, with a non-terminal reflecting its rank.
	By replacing the LMS inf-substrings by their \LMSOrder{}-ranks in the text while discarding the single character Lyndon factors, 
	we obtain the string $T^{(1)} := \texttt{EBDCACA}$,
	whose LMS inf-substrings are given on the right side of \cref{figExLMS}.
	Among these LMS inf-substrings, we only continue with \texttt{BDC} and \texttt{AC}.
	Since all LMS-inf substrings are distinct, 
	their \LMSOrder{}-ranks determine the \OmegaOrder{}-order of the \typeHoshi{} inf-suffixes as shown on the right side of \cref{figExRanking}.
	It is left to induce the \typeL{} and \typeS{} suffixes, which is done exactly as in the SAIS algorithm.
	We conduct these steps in \cref{figExInduce}, which finally lead us to \CSA{}.

	\begin{figure*}
	  \centering{\begin{tikzpicture}

\matrix[InductionMatrix] (sa) {1     & 2  & 3  & 4 & 5  & 6  & 7  & 8  & 9 & 10 & 11 & 12 & 13 & 14 & 15 & 16 & 17 & 18 & 19 & 20 & 3\\
  20    & 17 & 12 & 5 & 15 & 10 &    &    & 7 & 2  &    &    &    &    &    &    &    & 1  &    &    & 4\\
        &    &    &   &    &    & 19 & 14 &   &    &    &    & 9  & 18 & 13 & 6  & 4  &    & 16 & 11 & 5 \\
        &    &    &   &    &    &    &    &   &    &  8 & 3  &    &    &    &    &    &    &    &    & 6 \\
        &    &    &   &    &    &    &    &   &    &    &    &    &    &    &    &    &    &    &    & \\
  20 & 17 & 12 & 5  & 15 & 10 & 19 & 14 & 7  & 2 & 8 & 3  & 9  & 18 & 13 & 6 & 4 & 1  & 16 & 11 & 7 \\
  20 & 19 & 16 & 11 & 14 & 9  & 18 & 13 & 6 & 4 & 7 & 2 & 8 & 17 & 12  & 5 & 3 & 1 & 15  & 10  & 8\\
  a  & b  & d  & d  & b  & c  & c  & c  & c  & c & b & b  & b  & a  & a  & a & b & c  & a  & a  & {\tiny 9}\\
	};
	\node [left=0cm of sa-2-1] {\typeHoshi{} suffixes};
	\node [left=0cm of sa-3-1] {\typeL{} suffixes};
	\node [left=0cm of sa-4-1] {\typeS{} suffixes};
	\node [left=0cm of sa-8-1] {$\BBWT = $};
	\node [left=0cm of sa-7-1] {$\CSA-1= $};
	\node [left=0cm of sa-6-1] {$\CSA= $};

	\setcounter{PfeilCount}{1}
	\Pfeil{2}{2}{3}{7}{0.2em}
	\Pfeil{2}{3}{3}{19}{2.8em}
	\Pfeil{2}{4}{3}{20}{2em}
	\Pfeil{2}{5}{3}{8}{4em}
	\Pfeil{2}{6}{3}{13}{2.4em}
	\Pfeil{2}{9}{3}{16}{5.2em}
	\Pfeil{2}{10}{3}{17}{5.8em}
\Pfeil{3}{7}{3}{14}{4.5em}
	\Pfeil{3}{8}{3}{15}{4em}
\Pfeil{3}{17}{4}{12}{4em}
	\Pfeil{3}{13}{4}{11}{2em}

	\tikzstyle{Schrift} = [minimum height=1em, text height=0.5em, text depth=0.1pt, minimum width=2em, draw]

	\node [yshift=0.5em, fit=(sa-1-1.north) (sa-1-6.north), Schrift] (aS) {\typeS};
	\node [yshift=0.5em, fit=(sa-1-7.north) (sa-1-8.north), Schrift] (bL) {\typeL};
	\node [yshift=0.5em, fit=(sa-1-9.north) (sa-1-12.north),Schrift] (bS) {\typeS};
	\node [yshift=0.5em, fit=(sa-1-13.north) (sa-1-17.north),Schrift] (cL) {\typeL};
	\node [yshift=0.5em, fit=(sa-1-18.north),text width=1em, Schrift] (cS) {\typeS};
	\node [yshift=0.5em, fit=(sa-1-19.north) (sa-1-20.north),Schrift] (dL) {\typeL};

	\node [yshift=2em, fit=(sa-1-1.north) (sa-1-6.north), Schrift] (a) {\texttt{a}};
	\node [yshift=2em, fit=(sa-1-7.north) (sa-1-12.north), Schrift] {\texttt{b}};
	\node [yshift=2em, fit=(sa-1-13.north) (sa-1-18.north) (cS.south east), Schrift] {\texttt{c}};
	\node [yshift=2em, fit=(sa-1-19.north) (sa-1-20.north), Schrift] {\texttt{d}};

	\node [above=0em of sa-1-21] {\tiny{2}};
	\node [above=1.5em of sa-1-21] {\tiny{1}};

	\node [left=0cm of aS] {types};
	\node [left=0cm of a,align=left,yshift=1em] {starting\\character};

\end{tikzpicture}
	  }\caption{Inducing \typeL{} and \typeS{} inf-suffixes from the \LexOrder-order of the \typeHoshi{} inf-suffixes given in \cref{figExLMS}.
	    Rows~1 and~2 show the partitioning of $\CSA$ into buckets, first divided by the starting characters of the respective inf-suffixes, and second by the types \typeL{} and \typeS{}. 
	    Row~4 is $\CSA$ after inserting the \typeHoshi{} inf-suffixes according to their \LexOrder{}-order rank obtained from the right of \cref{figExRanking}.
	The \typeHoshi{} (resp.\ \typeL{}) inf-suffixes induce the \typeL{} (resp.\ \typeS{}) inf-suffixes in Row~5 (resp.\ Row~6).
	Putting all together yields \CSA{} in Row~7. 
	In the penultimate row $\CSA{}-1$, each text position stored in \CSA{} is decremented by one, wrapping around a Lyndon factor if necessary (for instance, $(\CSA{}-1)[2] = 19 = \iend{T_5}$ since $\CSA[2] = 17 = \ibeg{T_5}$).
	The last row shows $R[(\CSA - 1)[i]]$ in its $i$-th column, which is the BBWT of~$R$ as given in \cref{figComputingBBWT}.
      }
	  \label{figExInduce}

	\end{figure*}

\subsection{Correctness and Time Complexity}
Let us recall that our task is to compute the $\OmegaOrder$-order of the conjugates~$\conj[i_x-1]{T_x}$ for $i_x \in [1..|T_x|]$ 
of all Lyndon factors $T_1, \ldots, T_t$ of $R$.
We will frequently use that $\conj[i_x-1]{T_x} \OmegaOrder \conj[i_y-1]{T_y}$ is equivalent to $T_x[i_x..] \LexOrder T_y[i_y..]$ for $i_x \in [1..|T_x|]$ and $i_y \in [1..|T_y|]$.
We start with showing that the $\LMSOrder$-ranks of the LMS inf-substrings determine the $\LexOrder$-order of the \typeHoshi{} inf-suffixes\footnote{This is a counterpart to the property that 
the $\LMSOrder$-ranks of the LMS substrings determine the $\LexOrder$-order of the \typeHoshi{} suffixes~\cite[Theorem 3.12]{nong11sais}.},
whenever the LMS inf-suffixes are all distinct.

\begin{lemma}\label{lemLMSSubstringToSuffix}
  Let $S_x$ and $S_y$ be two LMS inf-substrings that are prefixes of $T_x[i_x..]$ and $T_y[i_y..]$, respectively,
  for $i_x \in [1..|T_x|]$ and $i_y \in [1..|T_y|]$.
  If $S_x \prec_{\textup{LMS}} S_y$ then
  $T_x[i_x..] \LexOrder T_y[i_y..]$.
\end{lemma}
\begin{proof}
  Given $S_x \prec_{\textup{LMS}} S_y$, 
  there is a position~$i$ such that (a) $S_x[i] < S_y[i]$ or (b) $S_x[i]$ is type~\typeL{} and $S_y[i]$ is type~\typeS{};
  let $i$ be the smallest such position.
  In the latter case (b), there is a position~$j > i$ 
  such that $T_x[i_x+j-1] = S_x[j] < S_x[i] = S_y[i] < S_y[j] = T_y[i_y+j-1]$ and $T_x[i_x..i_x+j-2] = T_y[i_y..i_y+j-2]$,
  where we abused the notation that $T_x[k] = (T_x T_x \cdots)[k]$ for a $k \in [1..2|T_x|]$.
  In both cases (a) and (b), $T_x[i_x..] \LexOrder T_x[i_y..]$.
\end{proof}

Exactly as in the SAIS recursion step, 
we map each LMS inf-substring to its respective meta-character via its $\LMSOrder$-rank, obtaining a string~$R^{(1)}$ whose characters are $\LMSOrder$-ranks.
The lexicographic order $\LexOrder$ induces a natural order on the strings whose characters are drawn from the $\LMSOrder$-ranks.
With that, we can determine the Lyndon factorization on $R^{(1)}$, which is given by the following connection:

\begin{lemma}\label{lemSameLyndonFactorization}
  There is a one-to-one correspondence between Lyndon factors of $R$ and $R^{(1)}$, 
  meaning that each Lyndon factor of $R^{(1)}$ generates a Lyndon factor in~$R$
  by expanding each of its $\LMSOrder$-ranks to the characters of the respective LMS inf-substring (while omitting the last character if it is the beginning of another LMS inf-substring),
  and vice-versa by contracting the characters of~$R$ to non-terminals.
\end{lemma}
\begin{proof}
  We first observe that each LMS inf-substring is contained in $T_x[1..|T_x|]T_x[1]$ for an $x \in [1..t]$.
  Now, let $L$ be a Lyndon factor of $R^{(1)}$ with $L = r_1 \cdots r_\ell$ such that each $r_i$ is a $\LMSOrder$-rank.
  Suppose that there is a $d \in [1..\ell-1]$ such that $r_1 \cdots r_d$ expands to a suffix~$T_x[s..|T_x|]$ of $T_x$ 
  (again omitting the last character of each expanded LMS inf-substring)
  and $r_{d+1} \cdots r_\ell$ expands to a prefix~$P$ of~$T_{x+1}$. 
  Since $L$ is a Lyndon word, $r_1 \cdots r_d \LexOrder r_1 \cdots r_\ell  \LexOrder r_{d+1} \cdots r_\ell$.
  Hence, $T_x[s..|T_x|] \LMSOrder T_x[s..|T_x|] T_x[1] \LMSOrder P$, 
  and with \cref{lemLMSSubstringToSuffix},
  $T_x[1..] \LexOrder T_{x}[s..] \LexOrder T_{x+1}[1..]$,
  contradicting the Lyndon factorization of~$R$ with \cref{lemLyndonOrderEquiv}.

  Finally, suppose that a Lyndon factor $L_1$ of $R^{(1)}$ expands to a proper prefix of a Lyndon factor~$T_x$.
  Let $L_2$ be its subsequent Lyndon factor, which has to end inside $T_x$ according to the above observation.
  Then $L_2 \LexOrder L_1$, which means that $T_x$ contains an inf-suffix smaller than $T_x$ due to \cref{lemLyndonOrderEquiv}, contradicting that $T_x$ is a Lyndon factor.
\end{proof}
Thanks to \cref{lemSameLyndonFactorization},
we do not have to compute the Lyndon factorization of~$R^{(1)}$ needed in the recursive step, but can infer it from the Lyndon factorization of~$R$.
Additionally, we have the property that the order of the LMS inf-substrings in the recursive step only depends on the Lyndon factors they are (originally) contained in.
It remains to show how the $\LMSOrder$-ranks of the LMS inf-substrings can be computed:

	\begin{figure*}
	  \centering{\begin{tikzpicture}

  \matrix[InductionMatrixSubstring] (sa) {1  & 2  & 3  & 4 & 5  & 6  & 7 & 8 & 9 & 10 & 11 & 12 & 13 & 14 & 15 & 16 & 17 & 18 & 19 & 20 & 3\\
 20  & 17 & 12 & 5 & 10 & 15 &   &   & 2 & 7  & 3  & 8  &    &    &    &    &    & 1  &    &    & 4\\
     &    &    &   &    &    & 19& 14&   &    &    &    & 9  & 18 & 13 &  4 &  6 &    & 16 & 11 & 5\\
     &    &    &   &    &    &   &   &   &    &    &    &    &    &    &    &    &    &    &    & \\
     & 17 & 12 & 5 & 15 & 10 &   &   & 7 & 2  & 8  & 3  &    &    &    &    &    &    &    &    & 6\\
     &    &    &   &    &    &   &   &   &    &    &    &    &    &    &    &    &    &    &    & \\
     & \texttt{A}   & \texttt{A}  & \texttt{B}  & \texttt{C}   & \texttt{C}   &  & & \texttt{D}  & \texttt{E}  &   &    &    &    &    &    &    &    &    &    &   7 \\
	};
	\node [left=0cm of sa-2-1] {\typeHoshi{} suffixes};
	\node [left=0cm of sa-3-1] {\typeL{} suffixes};
	\node [left=0cm of sa-5-1] {\typeS{} suffixes};
	\node [left=0cm of sa-7-1] {$\LMSOrder$-ranks};
\setcounter{PfeilCount}{1}

	\Pfeil{2}{2}{3}{7}{0.9em}
	\Pfeil{2}{3}{3}{19}{0.5em}
	\Pfeil{2}{4}{3}{20}{1.0em}
	\Pfeil{2}{5}{3}{13}{0.7em}
	\Pfeil{2}{6}{3}{8}{1.1em}

	\Pfeil{3}{7}{3}{14}{1.2em}
	\Pfeil{3}{8}{3}{15}{1.5em}
	\Pfeil{2}{9}{3}{16}{1.7em}
	\Pfeil{2}{10}{3}{17}{2em}
        
	\Pfeil{4}{20}{5}{6}{0.8em}
	\Pfeil{4}{19}{5}{5}{0.2em}
\Pfeil{4}{17}{5}{4}{0.4em}
	\Pfeil{4}{16}{5}{11}{0.4em}
	\Pfeil{4}{15}{5}{3}{0.5em}
	\Pfeil{4}{14}{5}{2}{0.6em}
	\Pfeil{4}{13}{5}{12}{0.7em}

	\Pfeil{5}{12}{5}{10}{1.5em}
	\Pfeil{5}{11}{5}{9}{1.3em}

	\tikzstyle{Schrift} = [minimum height=1em, text height=0.5em, text depth=0.1pt, minimum width=2em, draw]

	\node [yshift=0.5em, fit=(sa-1-1.north) (sa-1-6.north), Schrift] (aS) {\typeS};
	\node [yshift=0.5em, fit=(sa-1-7.north) (sa-1-8.north), Schrift] (bL) {\typeL};
	\node [yshift=0.5em, fit=(sa-1-9.north) (sa-1-12.north),Schrift] (bS) {\typeS};
	\node [yshift=0.5em, fit=(sa-1-13.north) (sa-1-17.north),Schrift] (cL) {\typeL};
	\node [yshift=0.5em, fit=(sa-1-18.north),text width=1em, Schrift] (cS) {\typeS};
	\node [yshift=0.5em, fit=(sa-1-19.north) (sa-1-20.north),Schrift] (dL) {\typeL};

	\node [yshift=2em, fit=(sa-1-1.north) (sa-1-6.north), Schrift] (a) {\texttt{a}};
	\node [yshift=2em, fit=(sa-1-7.north) (sa-1-12.north), Schrift] {\texttt{b}};
	\node [yshift=2em, fit=(sa-1-13.north) (sa-1-18.north) (cS.south east), Schrift] {\texttt{c}};
	\node [yshift=2em, fit=(sa-1-19.north) (sa-1-20.north), Schrift] {\texttt{d}};

	\node [above=0em of sa-1-21] {\tiny{2}};
	\node [above=1.5em of sa-1-21] {\tiny{1}};

	\node [left=0cm of aS] {types};
	\node [left=0cm of a,align=left,yshift=1em] {starting\\character};

\end{tikzpicture}
	  }\caption{Inducing LMS inf-substrings.
Thanks to the Lyndon factorization, we know the $\OmegaOrder$-order of the inf-suffixes starting with the Lyndon factors, which is
$T[20..] \OmegaOrder T[17..] \OmegaOrder T[12..] \OmegaOrder T[5..] \OmegaOrder T[2..] \OmegaOrder T[1..]$.
We insert the starting positions of these inf-suffixes in this order into their respective buckets, and fill the \typeHoshi{} buckets 
with the rest of \typeHoshi{} inf-suffixes by an arbitrary order (here we used the text order).
Like \cref{figInducingLMS}, the \typeHoshi{} (resp.\ \typeL{}) suffixes induce the \typeL{} (resp.\ \typeS{}) suffixes in Row~5 (resp.\ Row~6), but we skip those belonging to Lyndon factors of length one, since each of them is always stored at the leftmost position of its respective bucket.
In the last row, we assign each LMS inf-substring a non-terminal based on its \LMSOrder{}-rank, but omitting those that correspond to factors of length one.
	  }
	  \label{figInducingLMSinf}
	\end{figure*}
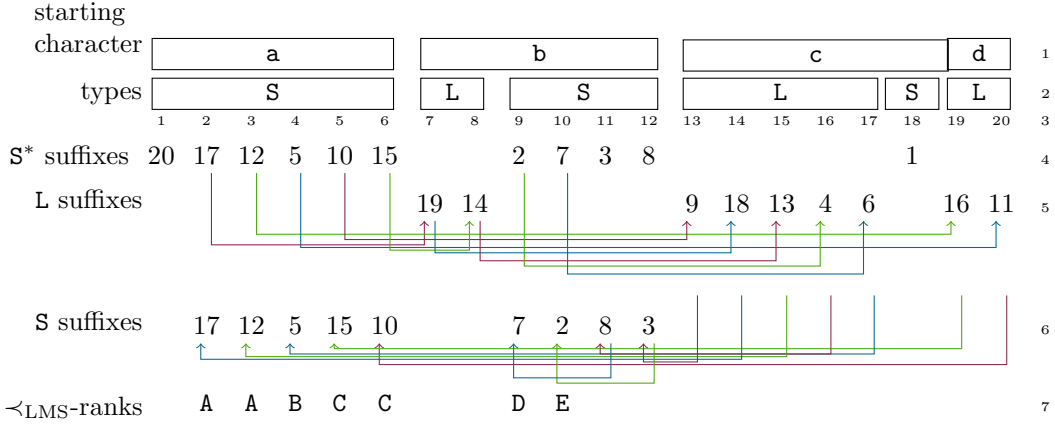

	\begin{lemma}\label{lemInduceLMSSubstrings}
  We can compute the $\LMSOrder$-ranks of all LMS inf-substrings in linear time.
\end{lemma}
\begin{proof}
  We follow the proof of \cite[Theorem 3.12]{nong11sais}.
  The idea is to know the $\LexOrder$-order among some smallest \typeHoshi{} inf-suffixes with which we can induce the $\LMSOrder$-ranks of all LMS inf-substrings.
  Here, we use the one-to-one correlation between each LMS inf-substring $R[i..j]$ and the respective \typeHoshi{} inf-suffix $R[i..]$ by using the starting position~$i$ for identification.
  To compute the order of the (traditional) LMS substrings, it sufficed to know the lexicographically smallest \typeHoshi{} suffix (cf.~\cref{figInducingLMS}), 
  which can be determined by appending an artificial character such as $\texttt{\$}$ to~$R$ with the property that it is smaller than all other characters appearing in~$R$.
  Here, we need to know the order of at least one \typeHoshi{} inf-suffix per Lyndon factor.
  That is because an inf-suffix can only induce the order of another inf-suffix of the \emph{same} Lyndon word.
  However, this is not a problem since we know that the inf-suffix starting with a Lyndon factor~$T_x$ is smaller in $\OmegaOrder$-order than all other inf-suffixes of~$T_x$, for each $x \in [1..t]$.
  In particular, we know that $T_{x} \LexOrderSucc T_{x+1}$ is equivalent to $T_{x} \OmegaOrderSucc T_{x+1}$ due to~\cite[Thm.~8]{bonomo14sorting},
  and hence we know the $\LexOrder$-ranks among all inf-suffixes starting with the Lyndon factors.\footnote{Since $T_t$ is the smallest Lyndon word, we have the invariants that $\CSA[1] = \ibeg{T_t}$ and $\BBWT[1] = R[\iend{T_t}] = R[|R|]$.}
  In what follows, we use the inf-suffixes starting with the Lyndon factors to induce the $\LMSOrder$-ranks of all LMS inf-substrings.

  However, the inducing only works if we include all text positions:
  While an ordered suffix $R[i..|R|]$ induces the order of $R[i-1..|R|]$ in the traditional SAIS, 
  here we want an inf-suffix $R[i..]$ to induce the order of $R[i-1..]$.
  For that, we define a superset of the LMS inf-substrings, whose elements are called LMS-prefixes~\cite[Sect.~3.4]{nong11sais}:
  Let $i \in [\ibeg{T_x}..\iend{T_x}]$ for an $x \in [1..t]$ be a text position, and
  let $j > i$ be the next \typeHoshi{} position in~$R$. 
  Then the \teigi{LMS-prefix}~$P_i$ starting at position~$i$ is
  $P_i := R[i..j]$ if $j \le \iend{T_x}$ or $P_i := R[i..j-1] \ibeg{T_x}$ if $j = \ibeg{T_{x+1}}$.
  In particular, if $i$ is the starting position of an LMS inf-substring~$S$, then $P_i = S$.
  The LMS-prefixes inherit the types (\typeL{} or \typeS{}) from their starting positions.
  We show that we can compute the $\LMSOrder$-ranks of all $P_i$'s by induce sorting:

  \block{Initialize the Suffix Array}
  We create \CSA{} of size~$|R|$ to store the $\LMSOrder$-ranks of all LMS-prefixes,
  where the entries are initially empty. 
  Like in SAIS, we divide \CSA{} into buckets, and put the LMS-prefixes corresponding to the LMS inf-substrings into the \typeS{} buckets of the respective starting characters 
  in lexicographically sorted order.
  See also \cref{figInducingLMSinf} for an example.

  \block{Inducing \typeL{} LMS-prefixes}
  We scan the suffix array from left to right, and take action whenever we access a non-empty value $i$ stored in $\CSA$:
  Given $i \in [\ibeg{T_x}..\iend{T_x}]$ and $i^- = i-1$ or $i^- = \iend{T_x}$ for $i = \ibeg{T_x}$,
  we insert $i^-$ into the \typeL{} bucket of the character $T_x[i^-]$ if $R[i^-..]$ is an \typeL{} inf-suffix.
  By doing so, we compute the $\LMSOrder{}$-order of all \typeL{} LMS-prefixes in ascending lexicographic order per \typeL{} bucket.
  The correctness follows by induction over the number~$k$ of inserted \typeL{} LMS-prefixes.
  Since we know that all LMS-prefixes~$P_{\ibeg{T_x}}$ for $x \in [1..t]$ starting with the Lyndon factors are stored correctly in $\LMSOrder$-order, and each of them is preceded by an \typeL{} LMS-prefix,
  we perform the insertion of the first \typeL{} LMS-prefix correctly, 
  which is induced by the lexicographically smallest \typeHoshi{} LMS-prefix $P_{T_t[1]}$.
  For the induction step, assume that there is a $k > 1$ such that when we append the $(k+1)$-th \typeL{} LMS-prefix $P_i$
  into its corresponding bucket, we have stored an \typeL{} LMS-prefix $P_j$ with larger $\LMSOrder$-rank in the same bucket.
  In this case, we have that $R[i] = R[j]$, $P_{j+1} \succ_{\textup{LMS}} P_{i+1}$ and $P_{j+1}$ is stored to the left of $P_{i+1}$.
  This implies that when we scanned \CSA{} from left to right, before appending $P_i$ to its bucket, we already did a mistake.

  The inducing step for the \typeS{} LMS-prefixes works exactly in the same way by symmetry.
Finally, we scan the computed \CSA{}, and for each pair of subsequent positions~$i$ and~$j$ with $i < j$ corresponding to the starting positions of two LMS inf-suffixes,
  we perform a character-wise comparison whether the LMS inf-substring starting at~$i$ is \LMSOrder{}-smaller than the one starting at~$j$.
  By doing so, we can compute the \LMSOrder{}-ranks of all LMS inf-substrings in linear time because the number of character comparisons is bounded by the number of characters covered by all LMS inf-substrings, which is \Oh{|R|}.
\end{proof}

With \cref{lemInduceLMSSubstrings}, we can determine the $\OmegaOrder{}$-order of the \typeHoshi{} inf-suffixes~$R$.
It is left to perform the induction step to induce first the order of the \typeL{} inf-suffixes, and subsequently the \typeS{} inf-suffixes,
which we do in the same manner as SAIS, but access $(T_x T_x \cdots )[i^-]$ instead of $R[i-1]$ when accessing a suffix array entry with value~$i$, 
where $x$ chosen such that $i \in [\ibeg{T_x}..\iend{T_x}]$ and $i^- = i-1$ or $i^- = \iend{T_x}$ if $i = \ibeg{T_x}$.
The correctness follows by construction:
Instead of partitioning the suffixes into LMS substrings
(maybe omitting a prefix of $R$ with \typeL{} suffixes),
we refine the Lyndon factors into a partitioning of LMS inf-substrings.

\block{Lyndon Factors of Length One}
It is left to reintroduce the Lyndon factors of lengths one to obtain the complete \CSA{} of $R$.
Remember that we omitted these factors at the recursive call.
After the recursive call, we reinsert each of them at the smallest position in the \typeS{} bucket of its respective starting character.
By doing so, we correctly sort them due to the following observation:
Suppose that there is a Lyndon factor consisting of a single character~$\texttt{b}$ (the following holds if $\texttt{b} \in \Sigma$ or if \texttt{b} is a rank of an LMS substring considered in the recursive call).
All LMS inf-substrings larger than one starting with \texttt{b} are larger than \texttt{bb} in the \OmegaOrder{}-order 
because such an LMS inf-substring starting with $R[i]$ having type~\typeHoshi{} is lexicographically smaller than~$R[i+1..]$.
Consequently, $\texttt{b}\texttt{b}\cdots \LexOrder R[i..] = \texttt{b} R[i+1..]$ since $\texttt{b}\cdots \LexOrder R[i+1..]$.
Thus, the Lyndon factor consisting of the single character~\texttt{b} does not have to be tracked further in the recursive call since we know that its rank precedes the ranks of all other LMS inf-substrings starting with \texttt{b}.

\block{Time Complexity}
By omitting Lyndon factors in the recursive calls, reducing $R$ to a string~$R'$ where 
no two subsequent inf-suffixes $R[i..]$ and $R[i+1..]$ are \typeHoshi{}, 
we can bound the maximum number of all \typeHoshi{} inf-suffixes by $n/2$ for the recursive call.
After the recursion, we can simply insert all omitted LMS inf-substrings into the order returned by the recursive call by a linear scan.
Hence, we obtain that $\mathcal{T}(n) = \mathcal{T}(n/2) + \Oh{n} = \Oh{n}$,
where $\mathcal{T}(n)$ is the time complexity for computing a circular suffix array of length~$n$.
Note that the omission of the single character Lyndon factors is crucial for obtaining this time complexity.
Without, there may be more than $n/2$ many \typeHoshi{} inf-suffixes,
and because we keep the same Lyndon factorization in all recursive levels, we could have $\Ot{n}$ LMS inf-suffixes at each recursion level.
The final step of computing the BBWT of~$T$ from the circular suffix array~$\CSA{}$ of~$R$ can be done in linear time with a linear scan of $\CSA{}$ as described in \cref{secComposedString}.

\subsection{Space Complexity}
Given that $z = \sum_{x=1}^t \tau_x$ is the number of all non-composed Lyndon factors~$F_1 \cdots F_z$,
the algorithm of \cref{lem:LyndonFactorizationLinearTime} computing the Lyndon factorization online only needs to maintain three integer variables of \Oh{\lg n} bits to find~$F_1 \cdots F_z$.
We can represent the non-composed Lyndon factorization by a bit vector~$B$ of length~$n$ marking the ending position of each factor $F_x$ ($x \in [1..z]$) with a one.
We additionally create a bit vector~$B_2$ of length~$z$, and mark the first occurrence of each non-composed Lyndon factor~$F_x$ in $B_2$ for $x \in [1..z]$ such that $B_2$ stores $t$ ones.
Then the $x$-th \bsq{1} in $B_2$ corresponds to the $x$-th composed Lyndon factor~$T_x$, and the number of \bsq{0}s between the $x$-th and $(x+1)$-th \bsq{1} in $B_2$ is $\tau_x-1$.
It is now possible to replace $T$ by $R$ and store the Lyndon factorization of~$R$ in $B$ (and resizing $B$ to length~$|R|$) since we can restore $T$ later with $B_2$.
(Alternatively, we can simulate $R$ having $T$ and $B_2$.)
This saves at least $(z-t) \lg \sigma \ge z - t $ bits, 
such that our working space is at most $n + t + n \lg \sigma$ bits including the text space, before starting the actual algorithm computing \CSA{}.
Building a rank-support data structure on~$B$ helps us to identify the Lyndon factor covering a text position of $R$ in constant time~\cite{jacobson89rank}.
A rank-support data structure provides support for a \teigi{rank query}, i.e., retrieving the number of ones up to a queried position in $B$.
Since a recursive call of SAIS works on a text instance of at most $|R|/2$ characters, 
we can rebuild $B$ from scratch by rerunning the algorithm of \cref{lem:LyndonFactorizationLinearTime} on $R^{(1)}$ or after finalizing the recursive call.
In total, we can maintain the Lyndon factorization in $n + \oh{n}$ bits with \Oh{n} total time throughout all recursive calls.
When a recursive call ends, we need to insert the omitted Lyndon factors of length one into the list of sorted \typeHoshi{} inf-suffixes.
But this can be done with a linear scan of the sorted \typeHoshi{} inf-suffixes and their initial characters, 
since we know that the omitted Lyndon factors have to be inserted at the first position among all inf-suffixes sharing the same initial character.
Additionally, we can achieve this within the space used for storing the circular suffix array~\CSA{}, since all \typeHoshi{} inf-suffixes use up at most half of the positions of the inf-suffix array.
Overall, we have an algorithm running with $n + t + \oh{n}$ bits on top of our modified SAIS, 
which uses $\Oh{\sigma \lg n}$ bits of working space additionally to \CSA{}.
If $\sigma$ is not constant, one may consider an option to get rid of this additional space requirement.
Luckily, we can do so with the in-place suffix array construction algorithm of {Goto}~\cite{goto19optimal} (or similarly with~\cite{li18sa}), which is a variation of SAIS,
storing an implicit representation of these $\Oh{\sigma \lg n}$ bits within the space of~\CSA{}.
Since $B_2$ is only needed for the final step computing the BBWT of $T$, 
we can compute \CSA{} with $n + \oh{n}$ additional bits of working space,
and \BBWT{} with $|\CSA| + n + t + \oh{n}$ additional bits of working space, where $|\CSA| = n \lg n$ denotes the size of $\CSA$ in bits.

\section{Conclusion}
We proposed an algorithm computing the bijective Burrows--Wheeler transform (BBWT) in linear time.
Consequently, we can also compute the extended Burrows--Wheeler transform (eBWT) within the same time bounds by a linear-time reduction of the problem to compute the eBWT to computing the BBWT\@.

Our trick was to first reduce our input text~$T$ to a text~$R$ by removing all duplicate Lyndon factors.
Second, we slightly modified the suffix array -- induce sorting (SAIS) algorithm to compute the $\OmegaOrder$-order of the conjugates of all Lyndon factors of~$R$ instead of the $\LexOrder$-order of all suffixes of~$R$.
For that, we introduced the notion of inf-suffixes and inf-substrings, and adapted the typing system of \typeL{}, \typeS{}, and \typeHoshi{} types from SAIS\@.
By some properties of the Lyndon factors, we could show that there are only some border cases, where a text position receives a different type in our modification.
Thanks to that, we could directly translate the induce sorting techniques of SAIS, 
and obtain the correctness of our result.

\subsection*{Open Problems}
The BBWT is bijective in the sense that it transforms a string of~$\Sigma^n$ into another string of $\Sigma^n$ while preserving distinctness.
Consequently, given a string of length~$n$, there is an integer~$k \ge 1$ with $\BBWT^{k}(T) = \BBWT^{k-1}(\BBWT(T)) = T$.
With our presented algorithm we can compute the smallest such number~$k$ in \Oh{nk} time.
However, we wonder whether we can compute this number faster, possible by scanning only the text in \Oh{n} time independent of~$k$.

We also wonder whether we can define the BBWT for the generalized Lyndon factorization~\cite{dolce19generalized}.
Contrary to the Lyndon factorization, the generalized Lyndon factorization uses a different order, called the \teigi{generalized lexicographic order}~$\GenOrder$.
In this order, two strings $S, T \in \Sigma^*$ are compared character-wise like in the lexicographic order.
However, the generalized lexicographic order~$\GenOrder$ can use different orders~$<_1, <_2, \ldots$ for each text position, i.e., 
$S \GenOrder T$ if and only if $S$ is a proper prefix of $T$ or
there is an integer~$\ell$ with $1 \le \ell \le \min(|S|,|T|)$ such that 
$S[1..\ell-1]=T[1..\ell-1]$ and $S[\ell] <_{\ell} T[\ell]$.

Recently, {Gibney and Thankachan}~\cite{gibney21finding} showed that finding an order of the alphabet such that the number of Lyndon factors of a given string is minimized or maximized is NP-complete.
This is an important but negative result for finding an advantage of the BBWT over the BWT, 
since the hope is to find a way to increase the number of Lyndon factors and therefore 
the chances of having multiple equal factors that are contracted to a single composed factor in the BBWT index of~\cite{bannai19bbwt}.
However, it is left open, whether we can find an efficient algorithm that approximates the alphabet order maximizing the number of Lyndon factor.

Another direction would be to find a string family for which we \CSA{} and \SA{} differ, for instance, with a relatively high Hamming distance.

\newpage

\bibliographystyle{plainurl}

\begin{thebibliography}{10}

\bibitem{adjeroh08bwt}
Donald Adjeroh, Timothy Bell, and Amar Mukherjee.
\newblock {\em The {Burrows--Wheeler} Transform: Data Compression, Suffix
  Arrays, and Pattern Matching}.
\newblock Springer, 2008.

\bibitem{bannai19bbwt}
Hideo Bannai, Juha K{\"{a}}rkk{\"{a}}inen, Dominik K{\"{o}}ppl, and Marcin
  Piatkowski.
\newblock Indexing the bijective {BWT}.
\newblock In {\em Proc.\ CPM}, volume 128 of {\em LIPIcs}, pages 17:1--17:14,
  2019.

\bibitem{bonomo14sorting}
Silvia Bonomo, Sabrina Mantaci, Antonio Restivo, Giovanna Rosone, and Marinella
  Sciortino.
\newblock Sorting conjugates and suffixes of words in a multiset.
\newblock {\em Int. J. Found. Comput. Sci.}, 25(8):1161, 2014.

\bibitem{burrows94bwt}
Michael Burrows and David~J. Wheeler.
\newblock A block sorting lossless data compression algorithm.
\newblock Technical Report 124, Digital Equipment Corporation, Palo Alto,
  California, 1994.

\bibitem{chen58lyndon}
Kuo~Tsai Chen, Ralph~H. Fox, and Roger~C. Lyndon.
\newblock Free differential calculus, {IV}. {T}he quotient groups of the lower
  central series.
\newblock {\em Annals of Mathematics}, pages 81--95, 1958.

\bibitem{dolce19generalized}
Francesco Dolce, Antonio Restivo, and Christophe Reutenauer.
\newblock On generalized {Lyndon} words.
\newblock {\em Theor. Comput. Sci.}, 777:232--242, 2019.

\bibitem{duval83lyndon}
Jean{-}Pierre Duval.
\newblock Factorizing words over an ordered alphabet.
\newblock {\em J. Algorithms}, 4(4):363--381, 1983.

\bibitem{ferragina00fmindex}
Paolo Ferragina and Giovanni Manzini.
\newblock Opportunistic data structures with applications.
\newblock In {\em Proc.\ FOCS}, pages 390--398, 2000.

\bibitem{ferragina05index}
Paolo Ferragina and Giovanni Manzini.
\newblock Indexing compressed text.
\newblock {\em J. {ACM}}, 52(4):552--581, 2005.

\bibitem{gagie18bwt}
Travis Gagie, Gonzalo Navarro, and Nicola Prezza.
\newblock Optimal-time text indexing in {BWT}-runs bounded space.
\newblock In {\em Proc.\ SODA}, pages 1459--1477, 2018.

\bibitem{gibney21finding}
Daniel Gibney and Sharma~V. Thankachan.
\newblock Finding an optimal alphabet ordering for {Lyndon} factorization is
  hard.
\newblock In {\em Proc.\ STACS}, volume 187 of {\em LIPIcs}, pages 35:1--35:15,
  2021.

\bibitem{gil12bbwt}
Joseph~Yossi Gil and David~Allen Scott.
\newblock A bijective string sorting transform.
\newblock {\em ArXiv 1201.3077}, 2012.
\newblock \href {http://arxiv.org/abs/1201.3077} {\path{arXiv:1201.3077}}.

\bibitem{goto19optimal}
Keisuke Goto.
\newblock Optimal time and space construction of suffix arrays and {LCP} arrays
  for integer alphabets.
\newblock In {\em Proc.\ PSC}, pages 111--125, 2019.

\bibitem{hon12ebwt}
Wing{-}Kai Hon, Tsung{-}Han Ku, Chen{-}Hua Lu, Rahul Shah, and Sharma~V.
  Thankachan.
\newblock Efficient algorithm for circular {Burrows--Wheeler} transform.
\newblock In {\em Proc.\ CPM}, volume 7354 of {\em LNCS}, pages 257--268, 2012.

\bibitem{hon11circular}
Wing{-}Kai Hon, Chen{-}Hua Lu, Rahul Shah, and Sharma~V. Thankachan.
\newblock Succinct indexes for circular patterns.
\newblock In {\em Proc.\ ISAAC}, volume 7074 of {\em LNCS}, pages 673--682,
  2011.

\bibitem{jacobson89rank}
Guy Jacobson.
\newblock Space-efficient static trees and graphs.
\newblock In {\em Proc.\ FOCS}, pages 549--554, 1989.

\bibitem{karkkainen06dc3}
Juha K{\"{a}}rkk{\"{a}}inen, Peter Sanders, and Stefan Burkhardt.
\newblock Linear work suffix array construction.
\newblock {\em J. {ACM}}, 53(6):918--936, 2006.

\bibitem{ko05sa}
Pang Ko and Srinivas Aluru.
\newblock Space efficient linear time construction of suffix arrays.
\newblock {\em J. Discrete Algorithms}, 3(2-4):143--156, 2005.

\bibitem{kufleitner09bwt}
Manfred Kufleitner.
\newblock On bijective variants of the {Burrows--Wheeler} transform.
\newblock In {\em Proc.\ PSC}, pages 65--79, 2009.

\bibitem{li18sa}
Zhize Li, Jian Li, and Hongwei Huo.
\newblock Optimal in-place suffix sorting.
\newblock In {\em Proc.\ SPIRE}, volume 11147 of {\em LNCS}, pages 268--284,
  2018.

\bibitem{lyndon54}
R.~C. Lyndon.
\newblock On {Burnside}'s problem.
\newblock {\em Transactions of the American Mathematical Society},
  77(2):202--215, 1954.

\bibitem{mantaci07ebwt}
Sabrina Mantaci, Antonio Restivo, Giovanna Rosone, and Marinella Sciortino.
\newblock An extension of the {Burrows--Wheeler} transform.
\newblock {\em Theor. Comput. Sci.}, 387(3):298--312, 2007.

\bibitem{nong11sais}
Ge~Nong, Sen Zhang, and Wai~Hong Chan.
\newblock Two efficient algorithms for linear time suffix array construction.
\newblock {\em {IEEE} Trans. Computers}, 60(10):1471--1484, 2011.

\end{thebibliography}

\end{document}